\documentclass[a4paper,11pt]{article}
\usepackage{jheppub} 
\usepackage{lineno}

\usepackage{bm}
\usepackage{mathrsfs}

\arxivnumber{2310.05610} 

\title{\boldmath Higher logarithms and $\varepsilon$-poles for the MS-like renormalization prescriptions}

\author[a]{N.P. Meshcheriakov,}
\author[b]{V.V. Shatalova}
\author[c,d,1]{and K.V. Stepanyantz\note{Corresponding author.}}

\affiliation[a]{Department of Quantum Theory and High Energy Physics, Moscow State University,\\Leninskie Gory, 119991 Moscow, Russia}
\affiliation[b]{AESC MSU – Kolmogorov boarding school, Moscow State University,\\Kremenchugskaya st. 11, 121357 Moscow, Russia}
\affiliation[c]{Department of Theoretical Physics, Moscow State University,\\Leninskie Gory, 119991 Moscow, Russia}
\affiliation[d]{Bogoliubov Laboratory of Theoretical Physics, JINR,\\Joliot-Curie st. 6, 141980 Dubna, Moscow region, Russia}

\emailAdd{np.meshcheryakov@physics.msu.ru}
\emailAdd{shatalova.viktoria@physics.msu.ru}
\emailAdd{stepan@m9com.ru}

\abstract{We consider a version of dimensional regularization (reduction) in which the dimensionful regularization parameter $\Lambda$ is in general different from the renormalization scale $\mu$. Then in the scheme analogous to the minimal subtraction the renormalization constants contain $\varepsilon$-poles, powers of $\ln\Lambda/\mu$, and mixed terms of the structure $\varepsilon^{-q}\ln^{p}\Lambda/\mu$. For the MS-like schemes we present explicit expressions for the coefficients at all these structures which relate them to the coefficients in the renormalization group functions, namely in the $\beta$-function and in the anomalous dimension. In particular, for the pure $\varepsilon$-poles we present explicit solutions of the 't~Hooft pole equations. Also we construct simple all-loop expressions for the renormalization constants (also written in terms of the renormalization group functions) which produce all $\varepsilon$-poles and logarithms and establish a number of relations between various coefficients at $\varepsilon$-poles and logarithms. The results are illustrated by some examples.}

\begin{document}
\maketitle
\flushbottom

\sloppy
\allowdisplaybreaks

\section{Introduction}
\hspace*{\parindent}

A regularization is an important ingredient needed for dealing with divergent expressions that appear in calculating quantum corrections for various quantum field theory models. A large number of various regularizations have been used in quantum field theory models, see \cite{Gnendiger:2017pys} for a review. The use of a proper regularization can allow revealing certain features of quantum corrections structure. For example, the higher covariant derivative regularization \cite{Slavnov:1971aw,Slavnov:1972sq,Slavnov:1977zf} in the supersymmetric version \cite{Krivoshchekov:1978xg,West:1985jx} (see also \cite{Aleshin:2016yvj,Kazantsev:2017fdc,Stepanyantz:2019lyo}) was the most important ingredient of the all-loop derivation of the NSVZ $\beta$-function in supersymmetric theories \cite{Novikov:1983uc,Jones:1983ip,Novikov:1985rd,Shifman:1986zi} made in \cite{Stepanyantz:2016gtk,Stepanyantz:2019ihw,Stepanyantz:2020uke}. It appeared \cite{Kataev:2013eta,Stepanyantz:2020uke} that the all-loop NSVZ renormalization scheme is given by the HD+MSL prescription \cite{Kataev:2013eta,Shakhmanov:2017wji,Stepanyantz:2017sqg} for which a theory is regularized by higher derivatives and minimal subtractions of logarithms are used for removing divergences. This implies that constructing the renormalization constants (for a theory in which all divergences are logarithmic) we include in them only powers of $\ln\Lambda/\mu$, where $\Lambda$ is a regularization parameter playing a role of an ultraviolet cutoff and $\mu$ is a renormalization point. All finite constants in this scheme are set to 0, so that for $\Lambda=\mu$ all renormalization constants are equal to 1 (or to the identity matrix). This prescription is certainly similar to the standard minimal subtraction \cite{tHooft:1973mfk} (or modified minimal subtraction \cite{Bardeen:1978yd}) which usually supplements the dimensional regularization \cite{'tHooft:1972fi,Bollini:1972ui,Ashmore:1972uj,Cicuta:1972jf} or reduction \cite{Siegel:1979wq}. In this case only $\varepsilon$-poles (where $\varepsilon\equiv 4-D$) are included into the renormalization constants. Evidently, $\ln\Lambda/\mu$ in the case of using the higher (covariant) derivative regularization (or other similar techniques) is analogous to $1/\varepsilon$ in the case of using the dimensional technique. It is well-known (see, e.g., \cite{Pronin:1997eb}) that in the one-loop approximation the coefficients at $1/\varepsilon$ and $\ln\Lambda/\mu$ are always the same. In higher orders the analogous relations are more complicated. For example, in $L$ loops the coefficient at $1/L\varepsilon$ coincides with the coefficient at $\ln\Lambda/\mu$ \cite{Chetyrkin:1980sa}. However, higher order divergent contributions to the effective action contain higher powers of $\varepsilon^{-1}$ in the case of using the dimensional regularization/reduction.
The coefficients at these higher poles satisfy the 't~Hooft pole equations \cite{tHooft:1973mfk} (see also \cite{Kazakov:2008tr} for a review). There are various generalizations of these equations to the different cases including even nonrenormalizable theories \cite{Kazakov:1987jp,Kazakov:2016wrp,Borlakov:2016mwp,Kazakov:2019wce,Solodukhin:2020vuw} and the analogous equations for logarithms in the renormalized Green functions \cite{Collins:1984xc}. Although (as far as we know) the explicit solutions of the 't~Hooft pole equations have not yet been constructed, these equations allow relating the coefficients at higher poles to the the coefficients of the renormalization group functions (RGFs), i.e. of the $\beta$-function and of the anomalous dimension.

Similarly, for theories regularized by higher derivatives divergences contain higher powers of logarithms. In the recent paper \cite{Meshcheriakov:2022tyi} (in the case of purely logarithmic divergences) the coefficients at all powers of logarithms present in the renormalization constants in the HD+MSL scheme were explicitly found in terms of the RGFs coefficients. However, it appears that in general it is not trivial to establish the correspondence between the functions which express the coefficients at higher $\varepsilon$-poles in terms of the RGFs coefficients (for theories regularized by the dimensional technique) and the similar functions giving the coefficients at higher logarithms (for theories regularized by higher derivatives). In this paper we will address this problem. For this purpose it is convenient to consider a version of the dimensional technique with two dimensionful parameters $\Lambda$ and $\mu$.\footnote{The usual dimensional technique is obtained in the particular case $\Lambda = \mu$.} The former one is the dimensionful parameter of the regularized theory, while the latter one is again the renormalization point. In this case divergences will contain both $\varepsilon$-poles and logarithms.\footnote{The parameter $\Lambda$ can in general be arbitrary. However, it is convenient to consider the limit $\Lambda \to \infty$ in order to establish the correspondence to the regularizations of the cut-off type. Therefore, it is reasonable to include the pure logarithms into the renormalization constants as well.} Certainly, mixed terms containing the products of logarithms and $\varepsilon$-poles also appear. Some explicit calculations made with the help of this technique can be found in \cite{Aleshin:2015qqc,Aleshin:2016rrr,Aleshin:2019yqj}. In the scheme analogous to the minimal subtraction ($\mbox{MS}$) prescription the renormalization constants contain only $\varepsilon$-poles, powers of $\ln\Lambda/\mu$ and the mixed terms. The modified minimal subtraction ($\overline{\mbox{MS}}$) scheme is obtained if the parameter $\Lambda$ is replaced by $\bar\Lambda = \Lambda  \exp(-\gamma/2)\sqrt{4\pi}$, where $\gamma\equiv - \Gamma'(1)\approx 0.577$. We will also consider the so-called $\mbox{MS}$-like schemes which (like the above mentioned $\overline{\mbox{MS}}$ scheme) differ from the $\mbox{MS}$ scheme by multiplying the parameter $\Lambda/\mu$ by a constant. Evidently, the analysis of terms with higher powers of logarithms and $\varepsilon$-poles made within the above described renormalization scheme can in particular establish the correspondence between the coefficients at $\varepsilon$-poles and logarithms. In this paper we present explicit expressions for all these coefficients (including the ones at the mixed terms) entering various renormalization constants in terms of the coefficients of the $\beta$-function and (for the matter field renormalization) the anomalous dimension. In particular, for pure $\varepsilon$-poles we present explicit solution of the 't~Hooft pole equations in the $\mbox{MS}$ scheme.

The paper is organized as follows. The dimensional technique with two dimensionful parameters is described in Section \ref{Section_Dimensional_Regularization}. The coefficients at all $\varepsilon$-poles, logarithms, and mixed terms in the expression $\ln Z_\alpha$, where $Z_\alpha$ is the charge renormalization constant, are found in Section \ref{Section_LnZ}. In this section we also present a simple expression for $\ln Z_\alpha$ which represents it explicitly via the $\beta$-function and produces all $\varepsilon$-poles and logarithms. Similar results for $(Z_\alpha)^S$, where $S$ is an arbitrary number, are obtained in Section \ref{Section_Z_Powers}. For the renormalization of fields the coefficients in $\ln Z$ (where $Z$ is the field renormalization constant) are constructed in Section \ref{Section_Field_Renormalization}. Again we present a simple expression for $\ln Z$ which relates it to the $\beta$-function and the anomalous dimension and produces all higher $\varepsilon$-poles and logarithms. Some relations between coefficients at higher $\varepsilon$-poles and logarithms are discussed in Section \ref{Section_Relations}. In particular, we discuss some interesting features in the structure of $\ln Z_\alpha$, $(Z_\alpha)^S$, and $\ln Z$. Some examples are considered in Section \ref{Section_Examples}. In particular, the three-loop expression for $\ln Z_\alpha$ and the two-loop expression for $\ln Z$ in ${\cal N}=1$ supersymmetric quantum electrodynamics (SQED) are verified in Section \ref{Subsection_N=1_SQED}. The five-loop expressions for
renormalization constants in a certain $\mbox{MS}$-like scheme (taken from \cite{Kleinert:2001ax}) for the $\varphi^4$-theory are compared with the general expressions derived in this paper in Sect. \ref{Subsection_Phi4}. The results are briefly summarized in Conclusion. Some explicit higher loop expressions for the renormalization constants are presented in Appendices.

\section{Dimensional technique with $\varepsilon$-poles and logarithms}
\label{Section_Dimensional_Regularization}

\subsection{Charge renormalization}
\hspace*{\parindent}\label{Subsection_Charge_Renormalization}

The most popular method for regularizing various quantum field theory models is dimensional regularization \cite{'tHooft:1972fi,Bollini:1972ui,Ashmore:1972uj,Cicuta:1972jf} or (in the supersymmetric case) dimensional reduction \cite{Siegel:1979wq}. In both cases the loop integrals are calculated in the non-integer dimension $D\equiv 4-\varepsilon$. This makes them convergent for $\varepsilon\ne 0$, and divergences correspond to the $\varepsilon$-poles. We will consider only renormalizable theories with a single dimensionless (in four space-time dimensions) coupling constant and a single mass parameter. (Certainly, it is possible to generalize our consideration to more complicated cases.) Note that the bare gauge coupling constant $\widetilde\alpha_0$ in the regularized theory has the dimension $m^{\varepsilon}$, so that it is standardly presented as

\begin{equation}\label{Renormalization_Standard}
\widetilde\alpha_0 = \mu^{\varepsilon} \bm{\alpha}\, \bm{Z_\alpha}^{-1}(\bm{\alpha}, 1/\varepsilon),
\end{equation}

\noindent
where $\mu$ is a renormalization point and $\bm{\alpha}$ is the (dimensionless) renormalized gauge coupling. The charge renormalization constant $\bm{Z_\alpha}$ absorbs divergences in the gauge part of the effective action. It contains $\varepsilon$-poles and some finite constants which determine a subtraction scheme. These finite constants are set to 0 for the simplest $\mbox{MS}$ renormalization prescription. However, it is more convenient to use the $\overline{\mbox{MS}}$ scheme \cite{Bardeen:1978yd}, when the parameter $\mu$ is replaced by the expression

\begin{equation}\label{MS_Bar_Mu}
\mu \to \frac{\mu\,\exp(\gamma/2)}{\sqrt{4\pi}}
\end{equation}

\noindent
and the renormalization constants again include only $\varepsilon$-poles. Note that the renormalization constants in the MS-like schemes are mass-independent \cite{tHooft:1973mfk,Collins:1973yy,Joglekar:1986be,Joglekar:1989ev}.

Although this technique is very convenient for making calculations, we will consider its modification \cite{Aleshin:2016rrr} which also contains logarithms similar to those that appear in the case of using the cut-off type regularizations (e.g., in the case of using the Slavnov's higher covariant derivative method \cite{Slavnov:1971aw,Slavnov:1972sq,Slavnov:1977zf}). For this purpose we present the bare coupling constant of a theory in $D$ dimensions in the form $\widetilde\alpha_0\equiv \Lambda^\varepsilon \alpha_0$, where $\Lambda$ is a dimensionful regularization constant analogous to the ultraviolet cut-off. To calculate the charge renormalization constant, one should first find the expression for the invariant charge. In the case of using the background field method it is obtained from the two-point Green function of the background gauge field and can be written as

\begin{equation}\label{Invariant_Charge}
d^{-1} = \Big(\frac{\Lambda}{P}\Big)^\varepsilon f\Big(\alpha_0 \Big(\frac{\Lambda}{P}\Big)^{\varepsilon},1/\varepsilon\Big) = \Big(\frac{\Lambda}{P}\Big)^\varepsilon \bigg[\frac{1}{\alpha_0} \Big(\frac{\Lambda}{P}\Big)^{-\varepsilon} + d_1(1/\varepsilon) + d_2(1/\varepsilon)\,\alpha_0 \Big(\frac{\Lambda}{P}\Big)^{\varepsilon} + \ldots \bigg],
\end{equation}

\noindent
where $P$ is the (absolute value of the Euclidean) momentum. The first term comes from the tree approximation, and the function $d_L(1/\varepsilon)$ corresponds to the $L$-loop approximation. The function $d_1$ is a polynomial of degree 1 in $1/\varepsilon$, while $d_L$ with $L\ge 2$ are polynomials in $1/\varepsilon$ of degree $L-1$. For simplicity, here we do not write down the mass dependence of the Green functions.

Written in terms of the renormalized (dimensionless) coupling constant and the normalization point $\mu$ the function in the left hand side of Eq. (\ref{Invariant_Charge}) should be finite in the limit $\varepsilon\to 0$, $\Lambda\to \infty$. After the replacement $\widetilde\alpha_0\to \Lambda^\varepsilon \alpha_0$ Eq. (\ref{Renormalization_Standard}) takes the form

\begin{equation}\label{Alpha_D_Renormalization}
\alpha_0 = \Big(\frac{\mu}{\Lambda}\Big)^\varepsilon \bm{\alpha} \bm{Z_\alpha}^{-1}(\bm{\alpha},1/\varepsilon).
\end{equation}

\noindent
Substituting this expression into Eq. (\ref{Invariant_Charge}) we present the invariant charge in the form

\begin{eqnarray}\label{Invariant_Charge_Renormalization}
&& d^{-1} = \Big(\frac{\Lambda}{P}\Big)^\varepsilon f\Big(\bm{\alpha}\, \bm{Z_\alpha}^{-1} \Big(\frac{\mu}{P}\Big)^{\varepsilon},1/\varepsilon\Big)
\nonumber\\
&&\qquad\qquad\qquad = \Big(\frac{\Lambda}{P}\Big)^\varepsilon \bigg[\frac{\bm{Z_\alpha}}{\bm{\alpha}} \Big(\frac{\mu}{P}\Big)^{-\varepsilon} + d_1(1/\varepsilon) + d_2(1/\varepsilon)\,\bm{\alpha} \bm{Z_\alpha}^{-1} \Big(\frac{\mu}{P}\Big)^{\varepsilon} + \ldots \bigg].\qquad
\end{eqnarray}

\noindent
The expression in the square brackets should be finite for all finite values of $P$ and, in particular, for $P=\mu$. Therefore, the function $\bm{Z_\alpha}(\bm{\alpha},1/\varepsilon)$ can be constructed from the requirement that the function $f(\bm{\alpha} \bm{Z_\alpha}^{-1},1/\varepsilon)$ be finite in the limit $\varepsilon\to 0$. Note that the renormalizability ensures that for $P\ne \mu$ the expression in the square brackets should also be finite. This can be achieved only if the coefficients at higher poles in the polynomials $d_L$ are related by certain equations to the coefficients at lower poles. Certainly, these equations should automatically be satisfied if the Feynman diagrams are calculated correctly.

Due to the finiteness of the expression in the square brackets in Eq. (\ref{Invariant_Charge_Renormalization}) the renormalized invariant charge is given by the expression

\begin{eqnarray}\label{Invariant_Charge_Renormalized}
&& d^{-1}\Big(\bm{\alpha},\ln\frac{\mu}{P}\Big) = \lim\limits_{\varepsilon\to 0} f\Big(\bm{\alpha} \bm{Z_\alpha}^{-1} \Big(\frac{\mu}{P}\Big)^{\varepsilon},1/\varepsilon\Big)\nonumber\\
&&\qquad\qquad\qquad = \lim\limits_{\varepsilon\to 0}  \bigg[\frac{\bm{Z_\alpha}}{\bm{\alpha}} \Big(\frac{\mu}{P}\Big)^{-\varepsilon} + d_1(1/\varepsilon) + d_2(1/\varepsilon)\,\bm{\alpha} \bm{Z_\alpha}^{-1} \Big(\frac{\mu}{P}\Big)^{\varepsilon}
+ \ldots \bigg].\qquad
\end{eqnarray}

Alternatively, the charge renormalization can be presented in the four-dimensional form

\begin{equation}\label{Alpha_Renormalization}
\frac{1}{\alpha_0} = \frac{Z_\alpha (\alpha, 1/\varepsilon, \ln \Lambda/\mu)}{\alpha},
\end{equation}

\noindent
where the function $Z_\alpha$ is a polynomial in $1/\varepsilon$ and $\ln\Lambda/\mu$. Namely, it contains $\varepsilon$-poles, logarithms, and the mixed terms, but {\it does not contain terms proportional to the positive powers of $\varepsilon$}. In the formalism under consideration the $\mbox{MS}$ renormalization constants will contain both $\varepsilon$-poles and $\ln\Lambda/\mu$, while all finite constants in them are set to 0. The modified minimal subtraction in this case corresponds to the renormalization prescription for which only various powers and products of $1/\varepsilon$ and $\ln\bar\Lambda/\mu$, where

\begin{equation}
\bar\Lambda \equiv \Lambda \exp(-\gamma/2)\sqrt{4\pi},
\end{equation}

\noindent
are admitted in the renormalization constants. Evidently, the standard dimensional technique is obtained in the particular case $\Lambda=\mu$, when all logarithms disappear. From the other side, the terms in $Z_\alpha$ without $\varepsilon$-poles look exactly like the renormalization constants for theories regularized by an ultraviolet cut-off, higher covariant derivative regularization, or another similar technique. In particular, the pure logarithmic terms in the MS or $\overline{\mbox{MS}}$ schemes (certainly for $\Lambda\ne \mu$ or $\bar\Lambda\ne\mu$, respectively) look like the renormalization constants in the HD+MSL scheme \cite{Kataev:2013eta,Shakhmanov:2017wji}.

The renormalization constant $Z_\alpha$ is also obtained from the finiteness of the invariant charge written in terms of the renormalized values,

\begin{eqnarray}\label{Invariant_Charge_Renormalized2}
&& d^{-1}\Big(\alpha,\ln\frac{\mu}{P}\Big) = \lim\limits_{\varepsilon\to 0} \Big(\frac{\Lambda}{P}\Big)^\varepsilon f\Big(\alpha Z_\alpha^{-1} \Big(\frac{\Lambda}{P}\Big)^{\varepsilon},1/\varepsilon\Big)\nonumber\\
&&\qquad\qquad\qquad
= \lim\limits_{\varepsilon\to 0} \Big(\frac{\Lambda}{P}\Big)^\varepsilon \bigg[\frac{Z_\alpha}{\alpha} \Big(\frac{\Lambda}{P}\Big)^{-\varepsilon} + d_1(1/\varepsilon) + d_2(1/\varepsilon)\,\alpha Z_\alpha^{-1} \Big(\frac{\Lambda}{P}\Big)^{\varepsilon} + \ldots \bigg].\qquad
\end{eqnarray}

Note that it is impossible to obtain the renormalization constant $Z_\alpha$ by naively comparing of Eqs. (\ref{Alpha_D_Renormalization}) and (\ref{Alpha_Renormalization}),

\begin{equation}
Z_\alpha(\alpha,1/\varepsilon,\ln\Lambda/\mu) \ne \Big(\frac{\Lambda}{\mu}\Big)^\varepsilon \bm{Z_\alpha}(\alpha,1/\varepsilon)\Big|_{\varepsilon^s\to 0\ \text{for all}\ s>0}.
\end{equation}

\noindent
(The condition ``$\varepsilon^s\to 0\ \text{for all}\ s>0$'' means that the terms proportional to the positive powers of $\varepsilon$ should be excluded from the considered expression.) Therefore, (for the same bare coupling constant) the renormalized coupling constants defined by Eqs. (\ref{Alpha_D_Renormalization}) and (\ref{Alpha_Renormalization}) are different. To distinguish them, we denote the former one by the bold font.

The relation between the renormalization constants $\bm{Z_\alpha}$ and $Z_\alpha$ can be constructed by comparing the renormalized invariant charges (\ref{Invariant_Charge_Renormalized}) and (\ref{Invariant_Charge_Renormalized2}). Then for $P=\Lambda$ we see that the relation between the functions  $\bm{Z_\alpha}(\alpha,1/\varepsilon)$ and $Z_\alpha(\alpha,1/\varepsilon,\ln\Lambda/\mu)$ can be written as

\begin{eqnarray}\label{Z_Relation}
&& \bigg[\frac{Z_\alpha}{\alpha} + d_1(1/\varepsilon) + d_2(1/\varepsilon)\,\alpha\, Z_\alpha^{-1} + \ldots \bigg] \nonumber\\
&&\qquad\qquad\qquad - \bigg[\frac{\bm{Z_\alpha}}{\alpha} \Big(\frac{\Lambda}{\mu}\Big)^{\varepsilon} + d_1(1/\varepsilon) + d_2(1/\varepsilon)\,\alpha\, \bm{Z_\alpha}^{-1} \Big(\frac{\mu}{\Lambda}\Big)^{\varepsilon}
+ \ldots \bigg] = O(\varepsilon),\qquad
\end{eqnarray}

\noindent
where $O(\varepsilon)$ denotes the terms which vanish in the limit $\varepsilon\to 0$. Note that the equality (\ref{Z_Relation}) is not trivial, because both square brackets contain the terms which depend on $\ln\Lambda/\mu$ and do not vanish in the limit $\varepsilon\to 0$. Taking into account that the functions $d_L$ contain $\varepsilon$-poles we see that the terms in $\bm{Z_\alpha} (\Lambda/\mu)^\varepsilon$ vanishing in the limit $\varepsilon\to 0$ contribute into the expression in the left hand side. Therefore, it is not so easy to find the renormalization constant $Z_\alpha(\alpha,1/\varepsilon,\ln\Lambda/\mu)$.

Setting $P=\mu$ in Eq. (\ref{Invariant_Charge_Renormalized}) we see that the expression $f(\bm{\alpha} \bm{Z_\alpha}^{-1}(\bm{\alpha},1/\varepsilon),1/\varepsilon)$ is finite in the limit $\varepsilon\to 0$ for any finite $\bm{\alpha}$. Certainly, it remains finite in this limit if we replace the coupling $\bm{\alpha}$ by the expression $\alpha(\Lambda/\mu)^\varepsilon$. This implies that

\begin{equation}
\bigg[\frac{1}{\alpha} \bm{Z_\alpha}\Big(\alpha\Big(\frac{\Lambda}{\mu}\Big)^\varepsilon, 1/\varepsilon\Big) \Big(\frac{\Lambda}{\mu}\Big)^{-\varepsilon} + d_1(1/\varepsilon) + d_2(1/\varepsilon)\,\alpha\, \bm{Z_\alpha}^{-1}\Big(\alpha\Big(\frac{\Lambda}{\mu}\Big)^\varepsilon, 1/\varepsilon\Big) \Big(\frac{\Lambda}{\mu}\Big)^{\varepsilon}
+ \ldots \bigg] = O(\varepsilon).
\end{equation}

\noindent
Therefore, looking at Eq. (\ref{Invariant_Charge_Renormalized2}) taken at $P=\mu$ it is tempting to identify the renormalization constant $Z_\alpha$ with $\bm{Z_\alpha}[\alpha(\Lambda/\mu)^\varepsilon, 1/\varepsilon]$. However, this is incorrect,

\begin{equation}\label{Naive_Renormalization}
Z_\alpha(\alpha,1/\varepsilon,\ln\Lambda/\mu) \ne \bm{Z_\alpha}\Big[\alpha \Big(\frac{\Lambda}{\mu}\Big)^\varepsilon,1/\varepsilon\Big],
\end{equation}

\noindent
because the right hand side contains the terms proportional to positive powers of $\varepsilon$, which are very essential. However, the $\varepsilon$-poles and finite constants evidently do not contain them. This implies that the terms without logarithms inside $Z_\alpha$ simply coincide with $\bm{Z_\alpha}$,

\begin{equation}\label{Z_Alpha_Equality}
Z_\alpha(\alpha,1/\varepsilon,\ln\Lambda/\mu)\Big|_{\mu=\Lambda} = Z_\alpha(\alpha,1/\varepsilon,0)= \bm{Z_\alpha}(\alpha,1/\varepsilon).
\end{equation}

\noindent
Moreover, taking into account that in the $\mbox{MS}$-like schemes (after a proper rescaling of $\Lambda/\mu$) the renormalization constant $\bm{Z_\alpha}(\alpha,1/\varepsilon)$ does not contain finite constants, we see that the terms with the first power of $\ln\Lambda/\mu$ in the right hand side of Eq. (\ref{Naive_Renormalization}) do not also contain positive powers of $\varepsilon$. Therefore, the terms without logarithms and the terms with the first power of $\ln\Lambda/\mu$ coincide in both sides of Eq. (\ref{Naive_Renormalization}). Differentiating both sides of Eq. (\ref{Naive_Renormalization}) with respect to $\ln\mu$ (at a fixed value of $\alpha$) using the chain rule in the right hand side and setting $\mu=\Lambda$ we obtain the relation

\begin{equation}\label{Z_Lowest_Relation}
\frac{\partial}{\partial\ln\mu} Z_\alpha(\alpha,1/\varepsilon,\ln\Lambda/\mu)\Big|_{\mu=\Lambda} = - \varepsilon\alpha \frac{\partial}{\partial\alpha} \bm{Z_\alpha}(\alpha,1/\varepsilon),
\end{equation}

\noindent
which should be valid in the $\mbox{MS}$-like schemes.

Divergences in the two-point Green function of the background gauge field can conveniently be encoded in the $\beta$-function. In the case of using the dimensional technique it is possible to introduce two different definitions for it. Namely, the $D$-dimensional $\beta$-function is defined by the equation

\begin{equation}\label{Beta_Definition_D}
\bm{\beta}(\bm{\alpha},\varepsilon) \equiv \frac{d\bm{\alpha}(\alpha_0(\Lambda/\mu)^\varepsilon,1/\varepsilon)}{d\ln\mu}\bigg|_{\alpha_0=\text{const}}
\end{equation}

\noindent
and certainly should not depend on both $\varepsilon$-poles and logarithms at a fixed value of the renormalized coupling constant $\bm{\alpha}$. Alternatively, one can introduce the {\it four-dimensional} $\beta$-function defined as

\begin{equation}\label{Beta_Definition}
\beta(\alpha) \equiv \frac{d \alpha(\alpha_0, 1/\varepsilon, \ln \Lambda/\mu) }{d \ln \mu}\bigg|_{\alpha_0=\text{const}},
\end{equation}

\noindent
which also depends on $\alpha_0$, $\ln\Lambda/\mu$, and $1/\varepsilon$ only via the renormalized coupling constant $\alpha$.

However, it is possible to find a simple relation between the $\beta$-functions (\ref{Beta_Definition_D}) and (\ref{Beta_Definition}) in the $\mbox{MS}$-like schemes. For this purpose we first consider Eq. (\ref{Alpha_D_Renormalization}) written in the form

\begin{equation}\label{D_Z_Definition}
\frac{1}{\alpha_0} \Big(\frac{\mu}{\Lambda}\Big)^\varepsilon = \frac{1}{\bm{\alpha}} \bm{Z_\alpha}
\end{equation}

\noindent
and differentiate it with respect to $\ln\mu$ at a fixed value of $\alpha_0$. Then after some simple transformations we obtain the equation

\begin{equation}
\frac{\varepsilon \bm{Z_\alpha}}{\bm{\alpha}} = \bm{\beta}(\bm{\alpha},\varepsilon) \frac{\partial}{\partial \bm{\alpha}}\Big(\frac{\bm{Z_\alpha}}{\bm{\alpha}}\Big),
\end{equation}

\noindent
which can equivalently be rewritten as

\begin{equation}\label{Beta_D_Diminations_Expression}
\bm{\beta}(\bm{\alpha},\varepsilon) = \varepsilon\bm{\alpha}\Big(-1 + \bm{\alpha}\frac{\partial\ln \bm{Z_\alpha}}{\partial\bm{\alpha}}\Big)^{-1}.
\end{equation}

From the other side, with the help of the chain rule for the derivative with respect to $\ln\mu$ the four-dimensional $\beta$-function can be presented in the form

\begin{equation}\label{Beta_New_Expression}
\beta(\alpha) = \alpha\frac{d\ln Z_\alpha}{d\ln\mu} = \alpha\frac{\partial\ln Z_\alpha}{\partial\alpha} \beta(\alpha) + \alpha\frac{\partial\ln Z_\alpha}{\partial\ln\mu},
\end{equation}

\noindent
where the total derivative $d/d\ln\mu$ is taken at $\alpha_0=\text{const}$ and acts on both explicit $\ln\mu$ and $\ln\mu$ inside the coupling constant $\alpha$. In contrast, the partial derivative $\partial/\partial\ln\mu$ acts only on the explicit $\ln\mu$. Note that this equation is valid for any value of $\mu$ and, in particular, for $\mu = \Lambda$. In this case the partial derivative with respect to $\ln\mu$ can be expressed from Eq. (\ref{Z_Lowest_Relation}). Moreover, the couplings $\alpha$ and $\bm{\alpha}$ evidently coincide for $\mu=\Lambda$. Therefore, it is possible to present the four-dimensional $\beta$-function in the form

\begin{equation}
\beta(\alpha) = \alpha\frac{\partial\ln \bm{Z_\alpha}}{\partial\alpha} \beta(\alpha) - \varepsilon\alpha^2 \frac{\partial \ln \bm{Z_\alpha}}{\partial\alpha}.
\end{equation}

\noindent
After adding $(\varepsilon\alpha-\beta(\alpha))$ to both sides of this equation it can equivalently be rewritten as

\begin{equation}\label{Auxiliary_Relation}
\varepsilon\alpha = \Big(\varepsilon\alpha -\beta(\alpha)\Big)\Big(1 - \alpha \frac{\partial \ln\bm{Z_\alpha}}{\partial\alpha}\Big).
\end{equation}

\noindent
From this equation and Eq. (\ref{Beta_D_Diminations_Expression}) (taken at $\mu=\Lambda$) we obtain the (well-known, see, e.g., \cite{Kazakov:2008tr}) relation between the $\beta$-functions in $D$ and $4$ dimensions,

\begin{equation}\label{Beta_Relation}
\bm\beta(\alpha,\varepsilon) = - \varepsilon\alpha + \beta(\alpha).
\end{equation}

Note that differentiating Eq. (\ref{D_Z_Definition}) with respect to $\ln\mu$ gives a similar equation

\begin{equation}
\bm{\beta}(\bm{\alpha},\varepsilon) = - \varepsilon\bm{\alpha} + \bm{\alpha} \frac{d\ln \bm{Z_\alpha}}{d\ln\mu}\bigg|_{\alpha_0=\text{const}},
\end{equation}

\noindent
so that in the $\mbox{MS}$-like schemes the four-dimensional $\beta$-function can be presented in two equivalent forms

\begin{eqnarray}
&& \beta\Big[\alpha\Big(\alpha_0,\frac{1}{\varepsilon}\Big)\Big] \equiv
\alpha \frac{d}{d\ln\mu} \ln Z_\alpha\Big[\alpha\Big(\alpha_0,\frac{1}{\varepsilon},\ln\frac{\Lambda}{\mu}\Big),\frac{1}{\varepsilon},\ln\frac{\Lambda}{\mu}\Big]\bigg|_{\mu=\Lambda}\nonumber\\
&&\qquad\qquad\qquad\qquad\qquad\qquad\qquad\quad = \alpha \frac{d}{d\ln\mu} \ln\bm{Z_\alpha}\Big[\bm{\alpha}\Big(\alpha_0\Big(\frac{\Lambda}{\mu}\Big)^\varepsilon, \frac{1}{\varepsilon}\Big),\frac{1}{\varepsilon}\Big]\bigg|_{\mu=\Lambda},\qquad
\end{eqnarray}

\noindent
where the derivative with respect to $\ln\mu$ is calculated at a fixed value of $\alpha_0$ and

\begin{equation}
\alpha\Big(\alpha_0,\frac{1}{\varepsilon}\Big) = \alpha\Big(\alpha_0,\frac{1}{\varepsilon},\ln\frac{\Lambda}{\mu}\Big)\bigg|_{\mu=\Lambda} = \bm{\alpha}\Big(\alpha_0\Big(\frac{\Lambda}{\mu}\Big)^\varepsilon, \frac{1}{\varepsilon}\Big)\bigg|_{\mu=\Lambda}.
\end{equation}

It is well-known (see, e.g., \cite{Collins:1984xc,Kazakov:2008tr}) that the $\beta$-function is given by the perturbative series

\begin{equation}\label{Beta_Perturbative_Expansion}
\beta(\alpha) = \sum\limits_{L=1}^\infty \beta_L \alpha^{L+1}.
\end{equation}

\noindent
The coefficients at various products of $\varepsilon$-poles and logarithms present in the renormalization constant $Z_\alpha$, namely at

\begin{equation}
\varepsilon^{-q} \ln^p \frac{\Lambda}{\mu},
\end{equation}

\noindent
with $p+q \ge 1$ can be expressed in terms of the coefficients $\beta_L$ in the $\mbox{MS}$-like schemes. In Sections \ref{Section_LnZ} and \ref{Section_Z_Powers} we construct the corresponding explicit expressions giving all coefficients in the expansions of $\ln Z_\alpha$ and $(Z_\alpha)^S$.

\subsection{Renormalization of fields}
\hspace*{\parindent}\label{Subsection_Field_Renormalization}

Next, we consider the renormalization of fields (or superfields). Let the corresponding two-point Green function is proportional to the function

\begin{equation}
G = G\Big[\alpha_0\Big(\frac{\Lambda}{P}\Big)^\varepsilon, 1/\varepsilon\Big] = 1 + g_1(1/\varepsilon)\, \alpha_0\Big(\frac{\Lambda}{P}\Big)^\varepsilon + g_2(1/\varepsilon)\, (\alpha_0)^2 \Big(\frac{\Lambda}{P}\Big)^{2\varepsilon} +\ldots,
\end{equation}

\noindent
where $g_L(1/\varepsilon)$ are polynomials in $1/\varepsilon$ of degree $L$. (As earlier, for simplicity, here we do not indicate the mass dependence of the function $G$.) The corresponding renormalized function $G_R$ is obtained by multiplying the function $G$ by the renormalization constant $\bm{Z}$. By definition, written in terms of the renormalized coupling constant $\bm{\alpha}$ the function $\bm{Z} G$ should be finite in the limit $\varepsilon\to 0$,

\begin{equation}\label{G_Renormalized}
G_R\Big(\bm{\alpha},\ln\frac{\mu}{P}\Big) = \lim\limits_{\varepsilon\to 0}\bm{Z}(\bm{\alpha},1/\varepsilon)\, G\Big[\Big(\frac{\mu}{P}\Big)^\varepsilon\bm{\alpha} \bm{Z_\alpha}^{-1}(\bm{\alpha},1/\varepsilon),1/\varepsilon\Big].
\end{equation}

\noindent
(Note that the renormalization constant $\bm{Z}$ depends only on $\bm{\alpha}$ and $1/\varepsilon$ and in the $\mbox{MS}$-like schemes does not depend on masses \cite{tHooft:1973mfk,Collins:1973yy,Joglekar:1986be,Joglekar:1989ev}.) In this formalism the anomalous dimension is defined by the equation

\begin{equation}\label{Gamma_Definition_D}
\bm{\gamma}(\bm{\alpha}) \equiv \frac{d\ln \bm{Z}(\bm{\alpha}, 1/\varepsilon)}{d\ln\mu} \bigg|_{\alpha_0=\text{const}} = \bm{\beta}(\bm{\alpha},\varepsilon) \frac{\partial \ln \bm{Z}}{\partial \bm{\alpha}},
\end{equation}

\noindent
where $\bm{\beta}(\bm{\alpha}, \varepsilon)$ has been introduced in Eq. (\ref{Beta_Definition_D}). Below we will see that the ($D$-dimensional) anomalous dimension (\ref{Gamma_Definition_D}) really depends only on $\bm{\alpha}$ and is independent of $\varepsilon$ (at least, in the $\mbox{MS}$-like schemes, which are considered in this paper).

Alternatively, one can construct the four-dimensional renormalization constant $Z(\alpha,1/\varepsilon,\ln\Lambda/\mu)$ by requiring the finiteness of the renormalized Green function

\begin{equation}\label{G_Renormalized_4D}
G_R\Big(\alpha,\ln\frac{\mu}{P}\Big) = \lim\limits_{\varepsilon\to 0} Z(\alpha,1/\varepsilon,\ln\Lambda/\mu)\, G\Big[\Big(\frac{\Lambda}{P}\Big)^\varepsilon\alpha Z_\alpha^{-1}(\alpha,1/\varepsilon,\ln\Lambda/\mu), 1/\varepsilon\Big].
\end{equation}

\noindent
Note that in this case the renormalization constant $Z$ depends not only on $1/\varepsilon$ and the four-dimensional renormalized coupling constant $\alpha$, but also on $\ln\Lambda/\mu$. Also, by definition, it cannot contain positive powers of $\varepsilon$. Then the four-dimensional anomalous dimension is defined by the equation

\begin{equation}\label{Gamma_Definition}
\gamma(\alpha) \equiv \frac{d\ln Z(\alpha,1/\varepsilon,\ln\Lambda/\mu)}{d\ln\mu}\bigg|_{\alpha_0=\text{const}},
\end{equation}

\noindent
in which the left hand side depends on $1/\varepsilon$ and $\ln\Lambda/\mu$ only through $\alpha(\alpha_0,1/\varepsilon,\ln\Lambda/\mu)$.

As in the case of the charge renormalization, the relation between the functions $\bm{Z}(\alpha,1/\varepsilon)$ and $Z(\alpha,1/\varepsilon,\ln\Lambda/\mu)$ is not trivial. Actually, it can be obtained by equating the expressions for the renormalized Green function $G_R$ (taken with the same argument $\alpha$). Setting $P=\Lambda$ in Eqs. (\ref{G_Renormalized}) and (\ref{G_Renormalized_4D}) we obtain the equation

\begin{eqnarray}\label{Z_Field_Relation}
&& Z\, \bigg[ 1 + g_1(1/\varepsilon)\, \alpha Z_\alpha^{-1} + g_2(1/\varepsilon)\, \alpha^2 Z_\alpha^{-2} +\ldots \bigg]\nonumber\\
&&\qquad\qquad - \bm{Z}\, \bigg[ 1 + g_1(1/\varepsilon)\, \alpha \bm{Z_\alpha}^{-1} \Big(\frac{\mu}{\Lambda}\Big)^\varepsilon + g_2(1/\varepsilon)\, \alpha^2 \bm{Z_\alpha}^{-2} \Big(\frac{\mu}{\Lambda}\Big)^{2\varepsilon} +\ldots \bigg]
= O(\varepsilon)\qquad
\end{eqnarray}

\noindent
analogous to Eq. (\ref{Z_Relation}). Note that each of two terms in the left hand side does not vanish in the limit $\varepsilon\to 0$ due to the dependence on $\ln\Lambda/\mu$, so that the above equation is nontrivial and really allows relating the renormalization constants $Z$ and $\bm{Z}$. However, using the finiteness of the expression $\bm{Z} G(\bm{\alpha}\bm{Z_\alpha}^{-1},1/\varepsilon)$ after the formal replacement $\bm{\alpha} \to \alpha(\Lambda/\mu)^\varepsilon$ we see that

\begin{equation}
\bm{Z}\Big(\alpha\Big(\frac{\Lambda}{\mu}\Big)^\varepsilon,1/\varepsilon\Big)\, G\Big[ \alpha \Big(\frac{\Lambda}{\mu}\Big)^\varepsilon \bm{Z_\alpha}^{-1}\Big(\alpha\Big(\frac{\Lambda}{\mu}\Big)^\varepsilon,1/\varepsilon\Big), 1/\varepsilon \Big] = O(\varepsilon).
\end{equation}

\noindent
Comparing this equation with Eq. (\ref{G_Renormalized_4D}) taken at $P=\mu$ it is tempting to identify naively the renormalization constant $Z_\alpha(\alpha,1/\varepsilon,\ln\Lambda/\mu)$ with $\bm{Z}(\alpha(\Lambda/\mu)^\varepsilon,1/\varepsilon)$. However, this is incorrect,

\begin{equation}\label{Naive_Idenitfication}
Z\Big[\alpha, 1/\varepsilon, \ln\frac{\Lambda}{\mu}\Big] \ne \bm{Z}\Big[\alpha\Big(\frac{\Lambda}{\mu}\Big)^\varepsilon,1/\varepsilon\Big],
\end{equation}

\noindent
because the right hand side contains important terms with positive powers of $\varepsilon$, which cannot be present in the four-dimensional renormalization constant $Z$. However, the pure $\varepsilon$-poles and the terms proportional to the first power of $\ln\Lambda/\mu$ in the $\mbox{MS}$ scheme evidently do not contain them. Therefore, these terms are the same in both sides of Eq. (\ref{Naive_Idenitfication}). This implies that the coefficients at pure $\varepsilon$-poles and at the terms with the first power of $\ln\Lambda/\mu$ in $Z$ and $\bm{Z}$ are related by the equations

\begin{eqnarray}\label{Z_Field_Equality}
&& Z(\alpha,1/\varepsilon,\ln\Lambda/\mu)\Big|_{\mu=\Lambda} = Z(\alpha,1/\varepsilon,0) = \bm{Z}(\alpha,1/\varepsilon);\\
\label{Z_Field_Lowest_Realtion}
&& \frac{\partial}{\partial\ln\mu} Z(\alpha,1/\varepsilon,\ln\Lambda/\mu)\Big|_{\mu=\Lambda} = -\varepsilon\alpha \frac{\partial}{\partial\alpha} \bm{Z}(\alpha,1/\varepsilon)
\end{eqnarray}

\noindent
analogous to Eqs. (\ref{Z_Alpha_Equality}) and (\ref{Z_Lowest_Relation}), respectively. Using them one can establish the correspondence between two definitions of the anomalous dimension presented above, namely, by Eqs. (\ref{Gamma_Definition_D}) and (\ref{Gamma_Definition}). For this purpose we use the chain rule for the derivative $d/d\ln\mu$ in Eq. (\ref{Gamma_Definition}), set $\mu=\Lambda$, and apply Eq. (\ref{Z_Field_Lowest_Realtion}). Then we obtain

\begin{equation}\label{Gamma_Auxiliary}
\gamma(\alpha) =  \Big(\beta(\alpha) \frac{\partial\ln Z}{\partial\alpha} + \frac{\partial\ln Z}{\partial\ln\mu}\Big)\bigg|_{\mu=\Lambda} = \big(\beta(\alpha) - \varepsilon\alpha\big) \frac{\partial\ln \bm{Z}}{\partial\alpha}\bigg|_{\mu=\Lambda}
= \bm{\beta}(\alpha,\varepsilon) \frac{\partial\ln \bm{Z}}{\partial\alpha},
\end{equation}

\noindent
where the last equality follows from Eq. (\ref{Beta_Relation}). According to Eq. (\ref{Gamma_Definition_D}), the expression in the right hand side is the anomalous dimension $\bm{\gamma}(\alpha)$. Therefore, we conclude that both definitions of the anomalous dimension give the same function,

\begin{equation}
\gamma(\alpha) = \bm{\gamma}(\alpha).
\end{equation}

\noindent
This in particular implies that the anomalous dimension $\bm{\gamma}$ does not (explicitly) depend on $\varepsilon$, because the anomalous dimension $\gamma$ does not depend on it.

It is well-known (see, e.g., \cite{Collins:1984xc,Kazakov:2008tr}) that the perturbative expansion of the anomalous dimension is written as

\begin{equation}\label{Gamma_Perturbative_Expansion}
\gamma(\alpha)=\sum\limits_{L=1}^\infty \gamma_L \alpha^{L},
\end{equation}

\noindent
where $\gamma_L$ corresponds to the $L$-loop contribution. Below in Section \ref{Section_Field_Renormalization} we will express all coefficients at various powers of $1/\varepsilon$ and $\ln\Lambda/\mu$ in the expansion of $\ln Z$ in terms of the coefficients $\gamma_L$ and $\beta_L$ in Eqs. (\ref{Beta_Perturbative_Expansion}) and (\ref{Gamma_Perturbative_Expansion}).

\section{Coefficients in the expansion of $\ln Z_\alpha$}
\hspace*{\parindent}\label{Section_LnZ}

Let us first express the coefficients at various products of $\varepsilon$-poles and logarithms in the expression $\ln Z_\alpha$, where $Z_\alpha$ is the charge renormalization constant defined by Eq. (\ref{Alpha_Renormalization}), in terms of the coefficients $\beta_n$ in Eq. (\ref{Beta_Perturbative_Expansion}). The perturbative expansion of $\ln Z_\alpha$ can be written in the form

\begin{equation}\label{LnZ_Expansion}
\ln Z_\alpha =  \sum\limits_{n=0}^\infty \sum\limits_{p=0}^\infty \sum\limits_{q=0}^\infty \alpha^{n+p+q} \widetilde{B}_{n+p+q,\,p,\,q} \, \varepsilon^{-q} \, \ln^p \frac{\Lambda}{\mu},
\end{equation}

\noindent
where $\widetilde B_{0,\,0,\,0} = 0$, and a number of loops corresponding to a certain term in this expression is equal to $L=n+p+q$. From this equation it is certainly evident that $L\geq p+q$. For the regularization under consideration the $\mbox{MS}$ scheme is defined by the condition

\begin{equation}
\widetilde B_{n,\,0,\,0} = 0, \quad n\ge 1,
\end{equation}

\noindent
which implies that all finite constants are set to 0, and only products of $\varepsilon$-poles and logarithms are included into the considered renormalization constant. From Eq. (\ref{LnZ_Expansion}) we see that the coefficients at pure poles and pure logarithms are given by $\widetilde B_{L,\,0,\,q}$ and $\widetilde B_{L,\,p,\,0}$, respectively.

In the $\mbox{MS}$-like schemes the ratio $\Lambda/\mu$ can differ from the one in the $\mbox{MS}$ scheme by a certain factor. After a proper redefinition of $\Lambda$ or $\mu$ it is possible to reduce the consideration of these schemes to the $\mbox{MS}$ case. That is why below we will discuss only the $\mbox{MS}$ renormalization prescription. Certainly, the results obtained in what follows are also valid in all $\mbox{MS}$-like schemes.

\subsection{Coefficients at pure poles}
\hspace*{\parindent}

First we find the coefficients $\widetilde B_{L,\,0,\,q}$ at pure poles. For this purpose we note that with the help of Eqs. (\ref{Z_Alpha_Equality}) and (\ref{Auxiliary_Relation}) it is possible to present the derivative of $\ln Z_\alpha$ with respect to $\ln\alpha$ in the form

\begin{equation}\label{LnZ_Pole_Expansion}
\frac{\partial\ln Z_\alpha}{\partial\ln\alpha}\bigg|_{\mu=\Lambda} = \frac{\beta(\alpha)}{\beta(\alpha)-\varepsilon\alpha}.
\end{equation}

\noindent
In the $\mbox{MS}$ scheme the terms with pure poles in Eq. (\ref{LnZ_Expansion}) can be written as

\begin{eqnarray}\label{lnZ_MS}
\ln Z_\alpha\Big|_{\mu=\Lambda} = \sum\limits_{n=0}^\infty \sum\limits_{q=1}^\infty \alpha^{n+q} \widetilde{B}_{n+q,\,0,\,q}\, \varepsilon^{-q}.
\end{eqnarray}

\noindent
(Note that now the index $q$ starts from 1, because, by definition, all finite constants, which corresponds to $q=0$, in the $\mbox{MS}$ scheme are set to 0.) From the other side, expanding the right hand side of Eq. (\ref{LnZ_Pole_Expansion}) into a series in $\varepsilon^{-1}$ we obtain

\begin{eqnarray}
\frac{\partial \ln Z_\alpha}{\partial \ln \alpha}\bigg|_{\mu=\Lambda} = -\frac{\beta(\alpha)/\varepsilon\alpha}{1 - \beta(\alpha)/\varepsilon\alpha} = - \sum\limits_{q=1}^\infty \Big( \dfrac{\beta(\alpha)}{\varepsilon \alpha} \Big)^q.
\end{eqnarray}

\noindent
After substituting in this equation the expansions (\ref{Beta_Perturbative_Expansion}) and (\ref{lnZ_MS}) we equate coefficients at the same powers of $\varepsilon$ and $\alpha$. This gives the values for the coefficients at pure poles

\begin{eqnarray}\label{Pole_Coefficients}
\widetilde{B}_{L,\,0,\,q} = - \frac{1}{L}  \sum\limits_{k_1, k_2, \ldots, k_q} \beta_{k_1}\beta_{k_2}\ldots\beta_{k_q} \bigg|_{\substack{k_1 + k_2 + \ldots+ k_q = L}},
\end{eqnarray}

\noindent
where $L\geq q$. The indices $q$, $k_1, \ldots, k_q$ range from 1 to infinity, the sum of all $k_i$ being equal to the number of loops $L$. In the particular case $q=1$ this equation relates the coefficients at the lowest ($q=1$) $\varepsilon$-poles in a certain loop to the corresponding contributions to the $\beta$-function,

\begin{equation}\label{B_L_0_1}
\beta_L = - L \widetilde{B}_{L,\,0,\,1}.
\end{equation}

\subsection{Coefficients at terms containing logarithms}
\hspace*{\parindent}

Next, it is necessary to find all coefficients at the terms containing logarithms. (They include both terms with pure logarithms and the mixed terms containing products of $\varepsilon$-poles and logarithms.) For this purpose we start with Eq. (\ref{Beta_New_Expression})

\begin{equation}\label{Main_Equation_For_lnZ}
\beta(\alpha) = \beta(\alpha)\,\frac{\partial \ln Z_\alpha}{\partial \ln \alpha}  + \alpha\, \frac{\partial \ln Z_\alpha}{\partial \ln \mu}.
\end{equation}

\noindent
Substituting the expression (\ref{LnZ_Expansion}) into this equation in the $\mbox{MS}$ scheme we obtain

\begin{eqnarray}\label{Main_Equation_Explicit_lnZ}
&&\beta(\alpha) =  \beta(\alpha)\,\sum\limits_{n=0}^\infty \sum\limits_{p=0}^\infty \sum\limits_{q=0}^\infty (n+p+q)\alpha^{n+p+q} \widetilde{B}_{n+p+q,\,p,\,q} \, \varepsilon^{-q} \, \ln^p \frac{\Lambda}{\mu}
\nonumber\\
&&\qquad\qquad\qquad\qquad\qquad\qquad - \sum\limits_{n=0}^\infty \sum\limits_{p=1}^\infty \sum\limits_{q=0}^\infty \, p \, \alpha^{n+p+q+1} \widetilde{B}_{n+p+q,\,p,\,q} \, \varepsilon^{-q} \, \ln^{p-1} \frac{\Lambda}{\mu}.\qquad
\end{eqnarray}

\noindent
After that, it is necessary to substitute here the perturbative expansion of the $\beta$-function given by Eq. (\ref{Beta_Perturbative_Expansion}) and equate the coefficients at $\alpha^{L+1} \varepsilon^0 \ln^0 \Lambda/\mu$. As a result we obtain the relation between the coefficients $\widetilde B_{L,\,1,\,0}$ (at the pure logarithms in the first power) and the corresponding contributions to the $\beta$-function,

\begin{equation}\label{B_L_1_0}
\beta_L = - \widetilde{B}_{L,\,1,\,0}.
\end{equation}

\noindent
Combining this equation with Eq. (\ref{B_L_0_1}) we see that the sum of the lowest poles and logarithms in the expression under consideration can be written as

\begin{equation}\label{LnZ_Lowest_Structure}
\ln Z_\alpha = -\sum\limits_{L=1}^\infty \alpha^L \beta_L \Big(\frac{1}{L\varepsilon} + \ln\frac{\Lambda}{\mu}\Big) + \mbox{higher poles and logarithms}
\end{equation}

\noindent
in agreement with \cite{Chetyrkin:1980sa}.

To find the coefficients in the remaining terms, we equate the coefficients at $\varepsilon^{-q} \ln^{p-1} \Lambda/\mu$  with $p\geq 1,q \geq 0$ and\footnote{The case $p=1$, $q=0$ should be considered separately, because for these values of $p$ and $q$ it is necessary to take into account the left hand side of Eq. (\ref{Main_Equation_Explicit_lnZ}).} $p+q\geq 2$ in Eq. (\ref{Main_Equation_Explicit_lnZ}). Multiplying the result by $1/\alpha p$ we obtain the recurrence relation

\begin{eqnarray}\label{Recursion_Relation}
&& \sum\limits_{n=0}^\infty \alpha^{n+p+q} \widetilde{B}_{n+p+q,\,p,\,q} = \frac{1}{p}\, \beta(\alpha) \sum\limits_{n=0}^\infty (n+p+q-1) \alpha^{n+p+q-2} \widetilde{B}_{n+p+q-1,\,p-1,\,q}
\nonumber\\
&& = \frac{1}{p}\,\beta(\alpha) \frac{d}{d\alpha}\, \sum\limits_{n=0}^\infty  \alpha^{n+p+q-1} \widetilde{B}_{n+p+q-1,\,p-1,\,q}
\end{eqnarray}

\noindent
because the original expression is written in terms of the same expression with $p\to p-1$. If $q \geq 1$, then repeating the process it is possible to relate it to the coefficients $\widetilde{B}_{n+q,\,0,\,q}$ (at pure $\varepsilon$-poles),\footnote{For $q=0$ it can be related to the coefficients $\widetilde B_{n,1,0}$ given by Eq. (\ref{B_L_1_0}). Below the result will be presented in a form which is also valid for this case.}

\begin{equation}\label{Recursion_Solution}
\sum\limits_{n=0}^\infty \alpha^{n+p+q} \widetilde{B}_{n+p+q,\,p,\,q} =  \frac{1}{p!} \Big(\beta(\alpha) \frac{d}{d\alpha} \Big)^{p} \sum\limits_{k=0}^\infty \alpha^{k+q} \widetilde{B}_{k+q,\,0,\,q}.
\end{equation}

\noindent
Substituting the perturbative expansion of the $\beta$-function (\ref{Beta_Perturbative_Expansion}) and equating the coefficients at the same powers of $\alpha$ we express the coefficients at higher mixed terms $\varepsilon^{-q} \ln^{p}\Lambda/\mu$ in terms of $\beta_n$ and the coefficients at pure poles,

\begin{eqnarray}\label{LnZ_Mixed_Terms_Result}
&&\hspace*{-5mm} \widetilde{B}_{n+p+q,\,p,\,q} = \frac{1}{p!} \, \sum\limits_{k=0}^\infty\,  \widetilde{B}_{k+q,\,0,\,q} \sum_{k_1=1}^\infty (q+k) \beta_{k_1} \sum_{k_2=1}^\infty (q+k+k_1) \beta_{k_2}\sum_{k_3=1}^\infty (q+k+k_1+k_2) \nonumber\\
&&\hspace*{-5mm} \times \beta_{k_3} \times \ldots \times \sum_{k_{p}=1}^\infty (q+k+k_1+k_2+\dots+k_{p-1})\beta_{k_{p}}\bigg|_{\substack{k+k_1+k_2+\dots +k_{p}=n+p}}.
\end{eqnarray}

\noindent
(In the case $p=1$ only the sums over $k$ and $k_1$ survive in this equation.) Eq. (\ref{LnZ_Mixed_Terms_Result}) is valid for all $n\geq 0$, $p\geq 1$, $q\geq 1$. Substituting into it the expression (\ref{Pole_Coefficients}) for the coefficients at pure $\varepsilon$-poles we can present the required coefficients in the form

\begin{equation}\label{LnZ_Result}
\widetilde B_{L,\,p,\,q} = -\frac{1}{L} \sum\limits_{k_1,k_2,\ldots, k_{p+q}} \beta_{k_1} \beta_{k_2} \ldots \beta_{k_{p+q}}\, \frac{K_{p+q}\bm{!}}{p!\, K_{q}\bm{!}}\,\bigg|_{K_{p+q}=L},
\end{equation}

\noindent
where we have introduced the notations

\begin{equation}\label{Generalized_Factorial}
K_m \equiv \sum\limits_{i=1}^m k_i; \qquad K_m\bm{!}\equiv K_1 K_2 \ldots K_{m};\qquad K_0\bm{!}\equiv 1.
\end{equation}

\noindent
The summation indices $k_1,\ldots, k_{p+q}$ range from 1 to $\infty$ and should satisfy the constraint $K_{p+q}=k_1+\ldots+k_{p+q} = L$.

Note that Eq. (\ref{LnZ_Result}) is valid for all $L\geq p+q \geq 1$, where $p,q\ge 0$. Really, for $p=0$ it produces the expression (\ref{Pole_Coefficients}), while for $q=0$ the coefficients at pure logarithms appear to be

\begin{eqnarray}\label{LnZ_Pure_Logarithms}
&& \widetilde B_{L,\,p,\,0} = -\frac{1}{p!} \sum\limits_{k_1,k_2,\ldots, k_{p}=1}^\infty \beta_{k_1} \beta_{k_2} \ldots \beta_{k_{p}} K_{p-1}\bm{!}\Big|_{K_{p}=L} = - \frac{1}{p!} \sum\limits_{k_1} \beta_{k_1} \sum\limits_{k_2} k_1 \beta_{k_2} \qquad\nonumber\\
&&\times  \sum\limits_{k_3} (k_1+k_2) \beta_{k_3} \times\ldots \times \sum\limits_{k_p} (k_1+k_2+\ldots + k_{p-1}) \beta_{k_p}\Big|_{k_1+k_2+\ldots + k_p=L}.
\end{eqnarray}

\noindent
This equation completely agrees with the expression for the coefficients in $\ln Z_\alpha$ in the HD+MSL scheme obtained in \cite{Meshcheriakov:2022tyi} if we take into account the difference of notations.\footnote{It is also necessary to remember that the $\beta$-function is scheme dependent starting from the three-loop approximation.} Namely, here in the right hand side of Eq. (\ref{LnZ_Expansion}) we write powers of the {\it renormalized} coupling constant $\alpha$, while in \cite{Meshcheriakov:2022tyi} the corresponding result contains powers of the {\it bare} coupling $\alpha_0$. If we rewrite the latter expression in terms of $\alpha$, then $\ln^p \Lambda/\mu$ will be replaced by $\ln^p \mu/\Lambda$ producing the multiplier $(-1)^p$.

\subsection{The result for $\ln Z_\alpha$}
\hspace*{\parindent}

Substituting the expression (\ref{LnZ_Result}) into the expansion (\ref{LnZ_Expansion}) we obtain the resulting expression for $\ln Z_\alpha$ in the form

\begin{equation}\label{lnZ_Final_Expansion}
\ln Z_\alpha = - \sum\limits_{\stackrel{\scriptsize \mbox{$p,q=0$}}{p+q\geq 1}}^\infty\, \sum\limits_{k_1,k_2,\ldots, k_{p+q}=1}^\infty \frac{1}{K_{p+q}}\cdot \frac{K_{p+q}\bm{!}}{p!\, K_q\bm{!}}\, \beta_{k_1} \beta_{k_2} \ldots \beta_{k_{p+q}}\, \alpha^{K_{p+q}}\, \varepsilon^{-q}\, \ln^p \frac{\Lambda}{\mu},
\end{equation}

\noindent
where $K_m\equiv k_1+k_2+\ldots + k_m;\ \ $ $K_m\bm{!}\equiv K_1 K_2\ldots K_m$, and $K_0\bm{!}\equiv 1$. Note that the term corresponding to $p=q=0$ in the expression (\ref{lnZ_Final_Expansion}) should be omitted because this case does not meet the condition $q+p\geq 1$. Evidently, the number of loops $L$ for a certain term is equal to $K_{p+q}$. The explicit expression for $\ln Z_\alpha$ in the five-loop approximation obtained from Eq. (\ref{lnZ_Final_Expansion}) is given by Eq. (\ref{LnZ_5Loops}) presented in Appendix \ref{Appendix_Explicit_Z}.

The expansion (\ref{lnZ_Final_Expansion}) can be encoded in a simple equation. To derive it we first differentiate Eq. (\ref{lnZ_Final_Expansion}) with respect to $\ln\alpha$ and rewrite the result in the form

\begin{eqnarray}\label{LnZ_Auxiliary}
&&\hspace*{-7mm} \frac{\partial\ln Z_\alpha}{\partial\ln\alpha} = - \sum\limits_{\stackrel{\scriptsize \mbox{$p,q=0$}}{p+q\geq 1}}^\infty\, \frac{1}{p!} \ln^p\frac{\Lambda}{\mu}\, \varepsilon^{-q} \sum\limits_{k_{q+p}=1}^\infty \beta_{k_{q+p}} \frac{\hat\partial}{\partial\ln\alpha} \alpha^{k_{q+p}} \sum\limits_{k_{q+p-1}=1}^\infty \beta_{k_{q+p-1}} \frac{\hat\partial}{\partial\ln\alpha}  \alpha^{k_{q+p-1}}
\nonumber\\
&&\hspace*{-7mm} \times \ldots \times \sum\limits_{k_{q+1}=1}^\infty \beta_{k_{q+1}} \frac{\hat\partial}{\partial\ln\alpha} \alpha^{k_{q+1}}
\sum\limits_{k_{q}=1}^\infty \beta_{k_{q}} \alpha^{k_{q}} \times \ldots\times \sum\limits_{k_1=1}^\infty \beta_{k_1} \alpha^{k_1},
\end{eqnarray}

\noindent
where we have introduced the differential operator $\hat\partial/\partial\ln\alpha$ which, by definition, acts on everything to the right of it. With the help of Eq. (\ref{Beta_Perturbative_Expansion}) this series can be presented as

\begin{equation}
\frac{\partial\ln Z_\alpha}{\partial\ln\alpha} = 1 - \sum\limits_{p=0}^\infty \frac{1}{p!} \Big(\ln\frac{\Lambda}{\mu}\, \frac{\hat\partial}{\partial\ln\alpha} \frac{\beta(\alpha)}{\alpha} \Big)^p\, \sum\limits_{q=0}^\infty\, \Big(\frac{\beta(\alpha)}{\varepsilon\alpha}\Big)^q.
\end{equation}

\noindent
Calculating the remaining sums over $p$ and $q$ we present the expression under consideration in the simple and beautiful form

\begin{equation}\label{LnZ_Final_Result}
\frac{\partial\ln Z_\alpha}{\partial\ln\alpha} = 1 - \exp\Big\{ \ln\frac{\Lambda}{\mu}\, \frac{\hat\partial}{\partial\ln\alpha} \frac{\beta(\alpha)}{\alpha}\Big\}\, \Big(1-\frac{\beta(\alpha)}{\varepsilon\alpha}\Big)^{-1}.
\end{equation}

\noindent
(Note that the differentiation with respect to $\ln\alpha$ in the left hand side is equivalent to multiplying each coefficient in $\ln Z_\alpha$ to the corresponding number of loops.) As a correctness check, we have also derived from Eq. (\ref{LnZ_Final_Result}) the five-loop expression (\ref{LnZ_5Loops}).

For completeness, here we also present the five-loop expressions for $\ln Z_\alpha$ in two particular cases. Namely, for pure $\varepsilon$-poles (the standard $\mbox{MS}$ ($\mbox{DR}$)-like schemes) Eq. (\ref{LnZ_5Loops}) gives

\begin{eqnarray}\label{LnZ_5Loops_Poles}
&&\hspace*{-4mm} \ln Z_\alpha\Big|_{\mu=\Lambda} = - \frac{\alpha \beta_1}{\varepsilon} -  \frac{\alpha^2}{2} \Big( \frac{\beta_2}{\varepsilon} + \frac{\beta_1^2}{\varepsilon^2} \Big)
- \frac{\alpha^3}{3} \Big( \frac{\beta_3}{\varepsilon} +  \frac{2 \beta_1\beta_2}{\varepsilon^2} + \frac{\beta_1^3}{\varepsilon^3} \Big)
- \frac{\alpha^4}{4} \Big( \frac{\beta_4}{\varepsilon} + \frac{2\beta_1\beta_3  +\beta_2^2}{\varepsilon^2}
\nonumber\\
&&\hspace*{-4mm} + \frac{3 \beta_1^2\beta_2}{\varepsilon^3} + \frac{\beta_1^4}{\varepsilon^4} \Big) - \frac{\alpha^5}{5} \Big(  \frac{\beta_5}{\varepsilon} + \frac{2(\beta_1\beta_4+\beta_2\beta_3)}{\varepsilon^2}
+ \frac{3\big(\beta_1^2 \beta_3 + \beta_1\beta_2^2\big)}{\varepsilon^3} +  \frac{4\beta_1^3\beta_2}{\varepsilon^4} + \frac{\beta_1^5}{\varepsilon^5} \Big) + O(\alpha^6),\nonumber\\
\end{eqnarray}

\noindent
while for pure logarithms the corresponding result is written as

\begin{eqnarray}\label{LnZ_5Loops_Logarithms}
&&\hspace*{-4mm} \ln Z_\alpha\Big|_{\varepsilon^{-1}\to 0} = - \alpha \beta_1 \ln\frac{\Lambda}{\mu} - \alpha^2 \Big(\beta_2 \ln\frac{\Lambda}{\mu} + \frac{\beta_1^2}{2} \ln^2\frac{\Lambda}{\mu} \Big)
- \alpha^3 \Big( \beta_3 \ln \frac{\Lambda}{\mu} +  \frac{3\beta_1\beta_2}{2} \ln^2\frac{\Lambda}{\mu}
\nonumber\\
&&\hspace*{-4mm} + \frac{\beta_1^3}{3} \ln^3\frac{\Lambda}{\mu} \Big) - \alpha^4 \Big( \beta_4 \ln \frac{\Lambda}{\mu} + \big(2\beta_1\beta_3  +\beta_2^2\big) \ln^2\frac{\Lambda}{\mu} + \frac{11 \beta_1^2\beta_2}{6} \ln^3\frac{\Lambda}{\mu} + \frac{\beta_1^4}{4} \ln^4\frac{\Lambda}{\mu}  \Big)
\nonumber\\
&&\hspace*{-4mm} - \alpha^5 \Big( \beta_5 \ln\frac{\Lambda}{\mu} + \frac{5}{2} (\beta_1\beta_4 + \beta_2\beta_3)\ln^2\frac{\Lambda}{\mu} + 3 \beta_1^2 \beta_3 \ln^3\frac{\Lambda}{\mu} + \frac{17\beta_1\beta_2^2}{6} \ln^3\frac{\Lambda}{\mu}
+ \frac{25\beta_1^3\beta_2}{12} \ln^4\frac{\Lambda}{\mu}
\nonumber\\
&&\hspace*{-4mm} + \frac{\beta_1^5}{5} \ln^5\frac{\Lambda}{\mu} \Big) + O(\alpha^6).
\end{eqnarray}

\noindent
(Up to notations) Eq. (\ref{LnZ_5Loops_Logarithms}) has exactly the same form as $\ln Z_\alpha$ in the HD+MSL scheme. However, it is necessary to remember that the coefficients $\beta_L$ are different in different renormalization schemes if $L\geq 3$ \cite{Vladimirov:1979ib,Vladimirov:1979my}.

\section{Coefficients in the expansion of $(Z_\alpha)^S$}
\hspace*{\parindent}\label{Section_Z_Powers}

Using the same method as in the previous section it is possible to find all coefficients in the expansion of the expression $(Z_\alpha)^S$, where $S$ is an arbitrary number. In our notation

\begin{equation}\label{ZS_Expansion}
(Z_\alpha)^S = 1 + \sum\limits_{n=0}^\infty \sum\limits_{p=0}^\infty \sum\limits_{q=0}^\infty \alpha^{n+p+q} B^{(S)}_{n+p+q,\,p,\,q}\, \varepsilon^{-q}\, \ln^p \frac{\Lambda}{\mu},
\end{equation}

\noindent
where $B^{(S)}_{0,\,0,\,0} = 0$ and in the $\mbox{MS}$ scheme $B^{(S)}_{n,\,0,\,0}=0$ for all $n\geq 1$. We will always assume this in what follows. Our purpose in this section is to express all coefficients $B^{(S)}_{n+p+q,\,p,\,q}$ in terms of the coefficients $\beta_L$ in the perturbative expansion of the $\beta$-function (\ref{Beta_Perturbative_Expansion}). Certainly, the results obtained below are also valid for all $\mbox{MS}$-like renormalization prescriptions.

\subsection{Coefficients at pure poles}
\hspace*{\parindent}

At the first step it is necessary to calculate the coefficients at pure poles in Eq. (\ref{ZS_Expansion}). For this purpose we consider Eq. (\ref{Alpha_D_Renormalization}) rewritten in the form

\begin{eqnarray}
\Big(\frac{\bm{\alpha}}{\alpha_0}\Big)^S = \Big(\frac{\Lambda}{\mu}\Big)^{\varepsilon S} \big(\bm{Z_\alpha}(\bm{\alpha}, 1/\varepsilon)\big)^S.
\end{eqnarray}

\noindent
Differentiating this equation with respect to $\ln\mu$ at a fixed value of $\alpha_0$ we obtain the relation

\begin{equation}
S \bm{\beta}(\bm{\alpha},\varepsilon) (\bm{Z_\alpha})^S = \Big(- S \varepsilon \bm{\alpha} + \bm{\beta}(\bm{\alpha},\varepsilon) \frac{\partial}{\partial \ln \bm{\alpha}}\Big) (\bm{Z_\alpha})^S.
\end{equation}

\noindent
After the formal replacement $\bm{\alpha} \to \alpha$ we rewrite it in terms of the four-dimensional $\beta$-function using Eq. (\ref{Beta_Relation}),

\begin{equation}
\beta(\alpha) \Big(\frac{\partial}{\partial \ln \alpha} -S \Big) (Z_\alpha)^S\Big|_{\mu=\Lambda}   = \varepsilon \alpha \frac{\partial (Z_\alpha)^S}{\partial \ln \alpha}\bigg|_{\mu=\Lambda},
\end{equation}

\noindent
where we also took Eq. (\ref{Z_Alpha_Equality}) into account. Next, we substitute into this equation the expansion

\begin{eqnarray}
(Z_\alpha)^S\Big|_{\mu=\Lambda} = 1+ \sum\limits_{n=0}^\infty \sum\limits_{q=1}^\infty \alpha^{n+q}\, B^{(S)}_{n+q,\,0,\,q}\,\varepsilon^{-q}
\end{eqnarray}

\noindent
and equate the coefficients at the same powers of $1/\varepsilon$ and $\alpha$. From the terms which do not contain $\varepsilon$-poles we obtain the coefficients

\begin{equation}\label{ZS_Lowest_Poles}
B^{(S)}_{L,\,0,\,1} = - \frac{S}{L} \beta_L
\end{equation}

\noindent
for all $L\geq 1$. Similarly, the terms with $\varepsilon$-poles give the recurrence relations for the remaining coefficients. The solution of these relations can be presented in the form

\begin{eqnarray}\label{ZS_Pure_Poles}
&& B^{(S)}_{L,\,0,\,q} =  - \frac{S}{L} \, \sum_{k_1}\, \beta_{k_1} \sum\limits_{k_2} \frac{(-S+K_1)}{K_1} \beta_{k_2} \sum\limits_{k_3} \frac{(-S+K_2)}{K_2} \beta_{k_3} \nonumber\\
&&\qquad\qquad\qquad\qquad\qquad\qquad\qquad\qquad\qquad \times \ldots \times
\sum_{k_q}\, \frac{(-S+K_{q-1})}{K_{q-1}} \beta_{k_q} \bigg|_{\substack{K_q=L}},\qquad
\end{eqnarray}

\noindent
where  $L\geq q\geq 2$ and the indices $k_i$ ranging from 1 to infinity should satisfy the constraint $K_q = k_1+k_2+\ldots+k_q = L$. Note that for the particular case $S=-1$ this result provides a solution for 't~Hooft pole equations in the MS scheme.

\subsection{Coefficients at terms containing logarithms}
\hspace*{\parindent}

Next, it is necessary to find coefficients at the terms containing logarithms (including the mixed terms) in the expansion (\ref{ZS_Expansion}). For this purpose we differentiate the equation

\begin{equation}
\Big(Z_\alpha (\alpha, 1/\varepsilon, \ln\Lambda/\mu)\Big)^S = \Big(\frac{\alpha}{\alpha_0}\Big)^S
\end{equation}

\noindent
(which follows from Eq. (\ref{Alpha_Renormalization})) with respect to $\ln\mu$ at a fixed value of $\alpha_0$. This gives the relation

\begin{equation}\label{ZS_Main_Equation}
\beta(\alpha) \Big(\frac{\partial}{\partial \ln \alpha} - S\Big) (Z_\alpha)^S + \alpha \frac{\partial (Z_\alpha)^S}{\partial\ln\mu} = 0.
\end{equation}

\noindent
We substitute the expansion (\ref{ZS_Expansion}) into this equation and equate the coefficients at the same powers of $1/\varepsilon$ and $\ln\Lambda/\mu$. After that, it is necessary to equate coefficients at the same powers of $\alpha$. In particular, equating the coefficients at $\alpha^{L+1} \varepsilon^0 \ln^0\Lambda/\mu$ we obtain

\begin{equation}\label{ZS_Lowest_Logarithms}
B^{(S)}_{L,\,1,\,0} = - S\beta_L.	
\end{equation}

\noindent
In combination with Eq. (\ref{ZS_Lowest_Poles}) this equation gives the expansion

\begin{equation}\label{ZS_Lowest_Structure}
(Z_\alpha)^S = 1 - S \sum\limits_{L=1}^\infty \alpha^L \beta_L \Big(\frac{1}{L\varepsilon} + \ln\frac{\Lambda}{\mu}\Big) + \mbox{higher poles and logarithms},
\end{equation}

\noindent
which evidently agrees with the analogous equation (\ref{LnZ_Lowest_Structure}) written for $\ln Z_\alpha$.

Similarly, equating the coefficients at $\alpha^{n+p+q+1} \varepsilon^{-q} \ln^{p-1}\Lambda/\mu$ we obtain the relation

\begin{eqnarray}
&&\hspace*{-5mm} B^{(S)}_{n+p+q,\,p,\,q} = \sum\limits_{n_1=0}^\infty B^{(S)}_{n_1+p-1+q,\,p-1,\,q} \sum\limits_{k_{q+p}=1}^\infty \frac{(-S+p-1+q+n_1)}{p} \beta_{k_{q+p}}  \bigg|_{n_1+k_{q+p}=n+1}\nonumber\\
&&\hspace*{-5mm} = \sum\limits_{n_2=0}^\infty B^{(S)}_{n_2+p-2+q,\,p-2,\,q} \sum\limits_{k_{q+p-1}=1}^\infty \frac{(-S+p-2+q+n_2)}{(p-1)} \beta_{k_{q+p-1}}\nonumber\\
&&\hspace*{-5mm} \qquad\qquad\  \times \sum\limits_{k_{q+p}=1}^\infty \frac{(-S+p-2+q+n_2+k_{q+p-1})}{p} \beta_{k_{q+p}} \bigg|_{n_2+k_{q+p-1}+k_{q+p}=n+2} = \ldots
\nonumber\\
\end{eqnarray}

\noindent
The second equality in this equation has been obtained by applying the first equality to the coefficient $B^{(S)}_{n_1+p-1+q,\,p-1,\,q}$. The summation index $n_2$ arising in this case is related to $n_1$ by the equation
$n_1 = n_2 + k_{q+p-1}-1$. Repeating the process it is possible to express the considered coefficients in terms of the coefficients at pure poles,

\begin{eqnarray}
&&\hspace*{-5mm} B^{(S)}_{n+p+q,\,p,\,q} = \frac{1}{p!}\, \sum\limits_{n_p=0}^\infty B^{(S)}_{n_p+q,\,0,\,q} \sum\limits_{k_{q+1}=1}^\infty (-S+q+n_p) \beta_{k_{q+1}} \sum\limits_{k_{q+2}=1}^\infty (-S+q+n_p+k_{q+1})  \nonumber\\
&&\hspace*{-5mm} \times \beta_{k_{q+2}} \times \ldots \times \sum\limits_{k_{q+p}=1}^\infty (-S+q+n_p + k_{q+1} +\ldots + k_{q+p-1}) \beta_{k_{q+p}} \Big|_{n_p+k_{q+1}+\ldots + k_{q+p}=n+p}.\nonumber\\
\end{eqnarray}

\noindent
Next, we substitute into this equation the expression (\ref{ZS_Pure_Poles}) for the coefficients at pure poles and obtain the result for the required coefficients in the form

\begin{equation}\label{ZS_Mixed_Terms_Result}
B^{(S)}_{L,\,p,\,q} = -\frac{S}{p!} \sum_{k_1,k_2,\ldots, k_{p+q}} \beta_{k_1} \beta_{k_2} \ldots \beta_{k_{p+q}}\, \frac{(-S+K_{p+q})\bm{!}}{K_q\bm{!}\, (-S+K_{p+q})}\, \bigg|_{K_{p+q}=L},
\end{equation}

\noindent
where we have introduced the notation

\begin{equation}\label{Generalized_Factorial_With_S}
(-S+K_m)\bm{!} \equiv (-S+K_1)(-S+K_2)\times \ldots \times (-S+K_m);\qquad (-S+K_0)\bm{!}\equiv 1,\ \
\end{equation}

\noindent
which for $S=0$ gives the generalized factorial defined by Eq. (\ref{Generalized_Factorial}). All indices $k_i$ in Eq. (\ref{ZS_Mixed_Terms_Result}) range from 1 to infinity and should satisfy the constraint $K_{p+q} = k_1+k_2+\ldots + k_{p+q} = L$. For $p=q=0$ the corresponding coefficients are equal to 0 for all $L\geq 1$ because we consider the $\mbox{MS}$ renormalization prescription.

The expression (\ref{ZS_Mixed_Terms_Result}) is valid for all $L\geq p+q\geq 1$, where $p,q\geq 0$. In particular, for $p=0$ it reproduces the expression (\ref{ZS_Pure_Poles}) for the coefficients at pure poles, while in the particular case  $q=0$ for the coefficients at pure logarithms we obtain the result

\begin{equation}\label{ZS_Logarithms_Result}
B^{(S)}_{L,\,p,\,0} = -\frac{S}{p!} \sum_{k_1,k_2,\ldots, k_{p}} \beta_{k_1} \beta_{k_2} \ldots \beta_{k_{p}}\, (-S+K_{p-1})\bm{!}\, \bigg|_{K_{p}=L},
\end{equation}

\noindent
which (up to notation) agrees with the one derived in \cite{Meshcheriakov:2022tyi} for the HD+MSL scheme under the assumption that only logarithmic divergences are present in a theory.

\subsection{The result for $(Z_\alpha)^S$}
\hspace*{\parindent}

After substituting the coefficients (\ref{ZS_Mixed_Terms_Result}) the expansion (\ref{ZS_Expansion}) takes the form

\begin{equation}\label{ZS_Final_Expansion}
(Z_\alpha)^S = 1 - S \sum\limits_{\stackrel{\scriptsize \mbox{$p,q=0$}}{p+q\geq 1}}^\infty\, \sum\limits_{k_1,k_2,\ldots, k_{p+q}=1}^\infty \frac{(-S+K_{p+q})\bm{!}}{p!\, K_q\bm{!}\, (-S+K_{p+q})}\, \beta_{k_1} \beta_{k_2} \ldots \beta_{k_{p+q}}\, \alpha^{K_{p+q}}\, \varepsilon^{-q}\, \ln^p \frac{\Lambda}{\mu},
\end{equation}

\noindent
where $K_m\equiv k_1+k_2+\ldots + k_m$, and the generalized factorial denoted by $\bm{!}$ is defined by Eqs. (\ref{Generalized_Factorial}) and (\ref{Generalized_Factorial_With_S}). Again, there are no terms for $p=q=0$, and the number of loops corresponding to a certain term is given by $K_{p+q}$.

As earlier, it is possible to construct a simple equation which encodes the expansion (\ref{ZS_Final_Expansion}), although it is less beautiful than the analogous Eq. (\ref{LnZ_Final_Result}) for $\ln Z_\alpha$. For this purpose we apply to both sides of Eq. (\ref{ZS_Final_Expansion}) the operator $\partial/\partial\ln\alpha -S$ and present the result in the form

\begin{eqnarray}\label{ZS_Auxiliary}
&& \Big(\frac{\partial}{\partial\ln\alpha} -S\Big) (Z_\alpha)^S = - S \sum\limits_{\stackrel{\scriptsize \mbox{$p,q=0$}}{p+q\geq 1}}^\infty\, \frac{1}{p!} \ln^p\frac{\Lambda}{\mu}\, \varepsilon^{-q} \sum\limits_{k_{q+p}=1}^\infty \beta_{k_{q+p}} \Big(\frac{\hat\partial}{\partial\ln\alpha} -S \Big)\alpha^{k_{q+p}}
\qquad\nonumber\\
&& \times \sum\limits_{k_{q+p-1}=1}^\infty \beta_{k_{q+p-1}} \Big(\frac{\hat\partial}{\partial\ln\alpha} - S\Big) \alpha^{k_{q+p-1}}\times \ldots \times \sum\limits_{k_{q+1}=1}^\infty \beta_{k_{q+1}} \Big(\frac{\hat\partial}{\partial\ln\alpha} - S\Big)\alpha^{k_{q+1}}
\qquad\nonumber\\
&& \times \sum\limits_{k_{q}=1}^\infty \beta_{k_{q}} \Big(1 - S \int\limits^{\wedge} \frac{d\alpha}{\alpha}\Big) \alpha^{k_{q}} \times \ldots\times  \sum\limits_{k_1=1}^\infty \beta_{k_1} \Big(1 - S \int\limits^{\wedge} \frac{d\alpha}{\alpha}\Big) \alpha^{k_1},
\end{eqnarray}

\noindent
where we introduced the operator $\int\limits^{\wedge} d\ln\alpha$ which acts {\it on everything to the right of it} according to the prescription

\begin{equation}
\int\limits^{\wedge} \frac{d\alpha}{\alpha}\,\alpha^n \equiv \frac{1}{n}\alpha^n
\end{equation}

\noindent
for $n>0$. (Evidently, only integer $n\geq 1$ can appear in Eq. (\ref{ZS_Auxiliary}).) Calculating the sums over $p$ and $q$ in Eq. (\ref{ZS_Auxiliary}) we rewrite the expression for $(Z_\alpha)^S$ in the form

\begin{equation}\label{ZS_Final_Result}
\Big(\frac{\partial}{\partial\ln\alpha} -S\Big) (Z_\alpha)^S = - S \exp\Big\{ \ln\frac{\Lambda}{\mu}\, \Big(\frac{\hat\partial}{\partial\ln\alpha} - S\Big) \frac{\beta(\alpha)}{\alpha}\Big\}\, \Big(1-\frac{\beta(\alpha)}{\varepsilon\alpha }+ S \int\limits^{\wedge} \frac{d\alpha}{\alpha}\, \frac{\beta(\alpha)}{\varepsilon\alpha} \Big)^{-1}.
\end{equation}

\noindent
(Certainly, this expression should be understood in the sense of the formal Taylor series expansion of the exponential function and of the fraction containing the integral operator.)

The five-loop expression for $(Z_\alpha)^S$ and the six-loop expression for $Z_\alpha$ derived from Eq. (\ref{ZS_Final_Expansion}) are presented in Appendix \ref{Appendix_Explicit_Z}, see Eqs. (\ref{ZS_5Loops}) and (\ref{Z_Alpha_6Loops}), respectively. Here we present only the six-loop expressions in the case $S=1$ (i.e. for $Z_\alpha$) for pure poles (corresponding to the standard $\mbox{MS}$ scheme)

\begin{eqnarray}\label{Z_Alpha_6Loops_Poles}
&&\hspace*{-1mm}  Z_\alpha\Big|_{\mu=\Lambda} = 1 - \frac{\alpha \beta_1}{\varepsilon} - \frac{\alpha^2 \beta_2}{2\varepsilon} - \alpha^3 \bigg[ \frac{\beta_3}{3\varepsilon} +\frac{\beta_1\beta_2}{6\varepsilon^2} \bigg]
- \alpha^4 \bigg[ \frac{\beta_4}{4\varepsilon} + \Big(\frac{\beta_1\beta_3}{6} + \frac{\beta_2^2}{8}\Big)\frac{1}{\varepsilon^2} + \frac{\beta_1^2\beta_2}{12\varepsilon^3}
\bigg]
\qquad\nonumber\\ 	
&&\hspace*{-1mm} - \alpha^5 \bigg[ \frac{\beta_5}{5\varepsilon} + \Big(\frac{3\beta_1\beta_4}{20} + \frac{7\beta_2\beta_3}{30}\Big)\frac{1}{\varepsilon^2} + \Big(\frac{\beta_1^2\beta_3}{10} + \frac{17\beta_1\beta_2^2}{120} \Big)\frac{1}{\varepsilon^3} + \frac{\beta_1^3\beta_2}{20\varepsilon^4}\bigg] - \alpha^6 \bigg[ \frac{\beta_6}{6\varepsilon} + \Big(\frac{2\beta_1\beta_5}{15}
\nonumber\\
&&\hspace*{-1mm} +\frac{5\beta_2\beta_4}{24} +\frac{\beta_3^2}{9}\Big) \frac{1}{\varepsilon^2} + \Big(\frac{\beta_1^2\beta_4}{10} + \frac{53\beta_1\beta_2\beta_3}{180} + \frac{\beta_2^3}{16}\Big) \frac{1}{\varepsilon^3} + \Big(\frac{\beta_1^3\beta_3}{15}
+ \frac{49\beta_1^2\beta_2^2}{360}\Big) \frac{1}{\varepsilon^4} + \frac{\beta_1^4\beta_2}{30\varepsilon^5}\bigg] \nonumber\\
&&\hspace*{-1mm} +\, O(\alpha^7)\vphantom{\bigg[}
\end{eqnarray}

\noindent
and for pure logarithms (corresponding to the HD+MSL scheme)

\begin{eqnarray}\label{Z_Alpha_6Loops_Logarithms}
&&\hspace*{-5mm}  Z_\alpha\Big|_{\varepsilon^{-1}\to 0} = 1 - \alpha \beta_1 \ln \frac{\Lambda}{\mu} - \alpha^2 \beta_2 \ln\frac{\Lambda}{\mu}
- \alpha^3 \bigg[ \beta_3 \ln\frac{\Lambda}{\mu} +\frac{\beta_1\beta_2}{2} \ln^2\frac{\Lambda}{\mu} \bigg]
- \alpha^4 \bigg[ \beta_4 \ln\frac{\Lambda}{\mu} \nonumber\\
&&\hspace*{-5mm} + \Big(\beta_1\beta_3 + \frac{\beta_2^2}{2}\Big) \ln^2 \frac{\Lambda}{\mu} + \frac{\beta_1^2\beta_2}{3} \ln^3\frac{\Lambda}{\mu} \bigg]
- \alpha^5 \bigg[ \beta_5 \ln\frac{\Lambda}{\mu} + \frac{3}{2}\big(\beta_1\beta_4 + \beta_2\beta_3\big) \ln^2\frac{\Lambda}{\mu}
\nonumber\\
&&\hspace*{-5mm} + \Big(\beta_1^2\beta_3 + \frac{5\beta_1\beta_2^2}{6}\Big) \ln^3 \frac{\Lambda}{\mu} + \frac{\beta_1^3\beta_2}{4} \ln^4\frac{\Lambda}{\mu}
\bigg]
- \alpha^6 \bigg[ \beta_6 \ln\frac{\Lambda}{\mu} + \big(2 \beta_1\beta_5 + 2 \beta_2\beta_4 + \beta_3^2\big) \ln^2\frac{\Lambda}{\mu}
\nonumber\\
&&\hspace*{-5mm} + \Big(2\beta_1^2\beta_4 + \frac{10\beta_1\beta_2\beta_3}{3} + \frac{\beta_2^3}{2}\Big) \ln^3\frac{\Lambda}{\mu} + \Big(\beta_1^3\beta_3 + \frac{13\beta_1^2\beta_2^2}{12}\Big) \ln^4\frac{\Lambda}{\mu}
+ \frac{\beta_1^4\beta_2}{5} \ln^5\frac{\Lambda}{\mu} \bigg] + O(\alpha^7),\nonumber\\
\end{eqnarray}

\noindent
where $\beta_L$ are scheme dependent starting from $L\geq 3$.

\section{Field renormalization constants}
\hspace*{\parindent}\label{Section_Field_Renormalization}

The logarithm of the field renormalization constant can be written in the form

\begin{equation}\label{LnZM_Expansion}
\ln Z = \sum\limits_{n=0}^\infty \sum\limits_{p=0}^\infty \sum\limits_{q=0}^\infty \alpha^{n+p+q} C_{n+p+q,\,p,\,q}\, \varepsilon^{-q}\, \ln^p\frac{\Lambda}{\mu},
\end{equation}

\noindent
where $C_{0,\,0,\,0}=0$. The $\mbox{MS}$ scheme corresponds to the case when $C_{L,\,0,\,0} = 0$ for all $L\geq 1$. In this section we express the coefficients $C_{n+p+q,\,p,\,q}$ in terms of the coefficients of the anomalous dimension and the $\beta$-function, see Eqs. (\ref{Beta_Perturbative_Expansion}) and (\ref{Gamma_Perturbative_Expansion}).

\subsection{Coefficients at pure poles}
\hspace*{\parindent}

As earlier, at the first step we calculate the coefficients at pure poles. By other words, we will find the expression for $\ln \bm{Z}$ in the standard MS scheme. We start from equation

\begin{equation}
\frac{\partial \ln Z}{\partial\ln\alpha}\bigg|_{\mu=\Lambda} = \frac{\alpha \gamma(\alpha)}{\beta(\alpha) - \varepsilon\alpha} = - \frac{\gamma(\alpha)}{\varepsilon} \Big(1 -  \frac{\beta(\alpha)}{\varepsilon \alpha} \Big)^{-1}
= - \frac{\gamma(\alpha)}{\varepsilon} \sum\limits_{n=0}^\infty \Big(  \frac{\beta(\alpha)}{\varepsilon \alpha} \Big)^{n},
\end{equation}

\noindent
which follows from Eq. (\ref{Gamma_Auxiliary}), and substitute into it the expansion for $\ln Z$ at $\mu=\Lambda$ following from Eq. (\ref{LnZM_Expansion}),

\begin{equation}
\ln Z\Big|_{\mu=\Lambda} = \sum\limits_{n=0}^\infty \sum\limits_{q=1}^\infty \alpha^{n+q} C_{n+q,\,0,\,q}\, \varepsilon^{-q}.
\end{equation}

\noindent
Equating the coefficients at the same powers of $\varepsilon$ and $\alpha$ we obtain

\begin{equation}\label{LnZM_Pure_Poles}
C_{L,\,0,\,q} = - \frac{1}{L} \sum_{k_1,k_2,\ldots, k_q} \gamma_{k_1} \beta_{k_2}\ldots \beta_{k_q} \Big|_{\substack{k_1+k_2+\ldots+ k_q=L}},
\end{equation}

\noindent
where  $L\geq q \geq 1$ and all $k_i$ are positive integers satisfying the constraint $k_1+k_2+\ldots+ k_q=L$. In particular, we see that the coefficients $C_{L,\,0,\,1}$ are related to the $L$-loop contribution to the anomalous dimension $\gamma_L$ by the equation

\begin{eqnarray}\label{LnZM_Lowest_Poles}
C_{L,\,0,\,1} = - \frac{1}{L} \gamma_{L}.
\end{eqnarray}

\subsection{Coefficients at terms containing logarithms}
\hspace*{\parindent}

The terms containing logarithms (including the mixed terms) can be found with the help of the equation

\begin{equation}\label{lnZM_Main_Equation}
\gamma(\alpha) \equiv \frac{d \ln Z (\alpha, 1/\varepsilon, \ln \Lambda/\mu)}{d \ln \mu} \Big|_{\alpha_0=\text{const}} = \beta(\alpha) \frac{\partial \ln Z}{\partial \alpha} +  \frac{\partial \ln Z}{\partial \ln \mu}.
\end{equation}

\noindent
Substituting into it the expansion (\ref{LnZM_Expansion}) and equating the coefficients at the terms proportional to $\alpha^L \varepsilon^0 \ln^0\Lambda/\mu$ we relate the coefficient $C_{L,\,1,\,0}$ to the $L$-loop contribution to the anomalous dimension,

\begin{equation}
C_{L,\,1,\,0} = -\gamma_L.
\end{equation}

\noindent
Combining this result with Eq. (\ref{LnZM_Lowest_Poles}) we obtain the equation analogous to Eqs. (\ref{LnZ_Lowest_Structure}) and (\ref{ZS_Lowest_Structure}),

\begin{equation}\label{LnZM_Lowest_Structure}
\ln Z = - \sum\limits_{L=1}^\infty \alpha^L \gamma_L \Big(\frac{1}{L\varepsilon} + \ln\frac{\Lambda}{\mu}\Big) + \mbox{higher poles and logarithms}.
\end{equation}

\noindent
Similarly, equating the coefficients at $\alpha^{n+p+q}\varepsilon^{-q} \ln^{p-1}\Lambda/\mu$ with $n\ge 0$, $p,q\geq 1$ we obtain the recurrence relation

\begin{equation}
C_{n+p+q,\,p,\,q} = \frac{1}{p} \sum\limits_{n_1=0}^\infty C_{n_1+p-1+q,\,p-1,\,q} \sum\limits_{k_{q+p}=1}^\infty (p-1+q+n_1) \beta_{k_{q+p}} \Big|_{n_1+k_{q+p}=n+1}.
\end{equation}

\noindent
It allows relating the coefficients in the left hand side to the coefficients at pure poles,

\begin{eqnarray}\label{LnZM_Recursion_Solution}
&&\hspace*{-5mm} C_{n+p+q,\,p,\,q} = \frac{1}{p!}\, \sum\limits_{n_p=0}^\infty C_{n_p+q,\,0,\,q} \sum\limits_{k_{q+1}=1}^\infty (q+n_p) \beta_{k_{q+1}} \sum\limits_{k_{q+2}=1}^\infty (q+n_p+k_{q+1})\beta_{k_{q+2}}\times \ldots \nonumber\\
&&\hspace*{-5mm} \times \sum\limits_{k_{q+p}=1}^\infty (q+n_p + k_{q+1} +\ldots + k_{q+p-1}) \beta_{k_{q+p}} \bigg|_{n_p+k_{q+1}+\ldots + k_{q+p}=n+p}.
\end{eqnarray}

\noindent
The coefficients $C_{n_{p+q},\,0,\,q}$ correspond to the pure $\varepsilon$-poles. The expression for them has been found earlier and is given by Eq. (\ref{LnZM_Pure_Poles}). Substituting it into Eq. (\ref{LnZM_Recursion_Solution}) we obtain the required coefficients in the form

\begin{equation}\label{LnZM_Result}
C_{L,\,p,\,q} = - \frac{1}{L}\sum\limits_{k_1,k_2,\ldots,k_{p+q}} \gamma_{k_1} \beta_{k_2} \beta_{k_3} \ldots \beta_{k_{p+q}}\, \frac{K_{p+q}\bm{!}}{p!\,K_q!}\bigg|_{K_{p+q}=L},
\end{equation}

\noindent
where the generalized factorial is defined by Eq. (\ref{Generalized_Factorial}). This equation is valid for all $L\geq p+q \geq 1$, where $p,q\geq 0$. In particular, for $p=0$ it reproduces Eq. (\ref{LnZM_Pure_Poles}) for the coefficients at pure poles, and for $q=0$ gives the coefficients at pure logarithms

\begin{eqnarray}
&& C_{L,\,p,\,0}  = - \sum\limits_{k_1,k_2,\ldots,k_{p}} \gamma_{k_1} \beta_{k_2} \beta_{k_3} \ldots \beta_{k_{p}}\, \frac{K_{p-1}\bm{!}}{p!}\bigg|_{K_{p}=L} = - \frac{1}{p!} \sum_{k_1} \gamma_{k_1}  \sum\limits_{k_2} k_1 \beta_{k_2} \qquad\nonumber\\
&&\times \sum\limits_{k_3} (k_1+k_2) \beta_{k_3} \times \ldots\times \sum\limits_{k_p} (k_1+k_2+\ldots+k_{p-1})\beta_{k_p}\bigg|_{\substack{k_1+k_2+\ldots+k_p=L}},
\end{eqnarray}

\noindent
which (up to notations) agree with the result obtained in \cite{Meshcheriakov:2022tyi}.

\subsection{The result for $\ln Z$}
\hspace*{\parindent}

The final result for $\ln Z$ obtained by substituting the coefficients (\ref{LnZM_Result}) into the expansion (\ref{LnZM_Expansion}) can be written as

\begin{equation}\label{lnZM_Final_Expansion}
\ln Z = - \sum\limits_{\stackrel{\scriptsize \mbox{$p,q=0$}}{p+q\geq 1}}^\infty\, \sum\limits_{k_1,k_2,\ldots, k_{p+q}=1}^\infty \frac{1}{K_{p+q}}\cdot \frac{K_{p+q}\bm{!}}{p!\, K_q\bm{!}}\, \gamma_{k_1} \beta_{k_2} \beta_{k_3} \ldots \beta_{k_{p+q}}\, \alpha^{K_{p+q}}\, \varepsilon^{-q}\, \ln^p \frac{\Lambda}{\mu},
\end{equation}

\noindent
where $K_m\equiv k_1+k_2+\ldots +k_m$, and $K_m\bm{!}$ is defined by Eq. (\ref{Generalized_Factorial}). The terms with $p=q=0$ are absent, and in each term the number of loops $L$ is equal to $K_{p+q}$. The explicit five-loop expression for $\ln Z$ is given by Eq. (\ref{LnZM_5Loops}) in Appendix \ref{Appendix_Explicit_Z}. Note that after the formal replacement $\gamma_L \to \beta_L$ this expression gives the corresponding result (\ref{lnZ_Final_Expansion}) for $\ln Z_\alpha$.

Differentiating Eq. (\ref{lnZM_Final_Expansion}) with respect to $\ln\alpha$ we can obtain an equation analogous to Eq. (\ref{LnZ_Auxiliary}). However, in this case it is necessary to consider terms with $q\geq 1$ and $q=0$ separately,

\begin{eqnarray}
&&\hspace*{-5mm} \frac{\partial\ln Z}{\partial\ln\alpha} = - \sum\limits_{p=0}^\infty \frac{1}{p!} \ln^p\frac{\Lambda}{\mu}\, \bigg\{\sum\limits_{q=1}^\infty\, \varepsilon^{-q} \sum\limits_{k_{q+p}}\beta_{k_{q+p}} \frac{\hat\partial}{\partial\ln\alpha} \alpha^{k_{q+p}} \sum\limits_{k_{q+p-1}}\beta_{k_{q+p-1}} \frac{\hat\partial}{\partial\ln\alpha} \alpha^{k_{q+p-1}}
\nonumber\\
&&\hspace*{-5mm} \times  \ldots\times \sum\limits_{k_{q+1}} \beta_{k_{q+1}} \frac{\hat\partial}{\partial\ln\alpha} \alpha^{k_{q+1}}
\sum\limits_{k_{q}} \beta_{k_{q}} \alpha^{k_{q}} \times \ldots \times \sum\limits_{k_2} \beta_{k_2} \alpha^{k_2}
+ \sum\limits_{k_{p}}\beta_{k_{p}} \frac{\hat\partial}{\partial\ln\alpha} \alpha^{k_{p}}\nonumber\\
&&\hspace*{-5mm} \times \sum\limits_{k_{p-1}}\beta_{k_{p-1}} \frac{\hat\partial}{\partial\ln\alpha} \alpha^{k_{p-1}} \times \ldots \times
\sum\limits_{k_{2}} \beta_{k_{2}} \frac{\hat\partial}{\partial\ln\alpha} \alpha^{k_2} \cdot\frac{\hat\partial}{\partial\ln\alpha}
\bigg\} \sum\limits_{k_1} \gamma_{k_1} \alpha^{k_1},\qquad
\end{eqnarray}

\noindent
where the sums over all $k_i$ are taken from 1 to infinity. Note that for $p=0$ the first of these sums in the first term is over $k_q$, and there are no sums in the last term. For $q=1$ only one sum without the operator $\hat\partial/\partial\ln\alpha$ is present in the first term. Using the perturbative expansions of the $\beta$-function and the anomalous dimension (Eqs. (\ref{Beta_Perturbative_Expansion}) and (\ref{Gamma_Perturbative_Expansion}), respectively) this expression can be rewritten as

\begin{eqnarray}
&& \frac{\partial\ln Z}{\partial\ln\alpha} = - \sum\limits_{p=0}^\infty \frac{1}{p!} \Big(\ln\frac{\Lambda}{\mu}\, \frac{\hat\partial}{\partial\ln\alpha} \frac{\beta(\alpha)}{\alpha} \Big)^p\, \sum\limits_{q=1}^\infty\, \Big(\frac{\beta(\alpha)}{\varepsilon\alpha}\Big)^{q-1} \frac{\gamma(\alpha)}{\varepsilon}\nonumber\\
&&\qquad\qquad\qquad\qquad\qquad\qquad\qquad\qquad\quad - \sum\limits_{p=1}^\infty \frac{1}{p!} \Big(\ln\frac{\Lambda}{\mu}\, \frac{\hat\partial}{\partial\ln\alpha} \frac{\beta(\alpha)}{\alpha} \Big)^p\, \frac{\alpha\gamma(\alpha)}{\beta(\alpha)}.\qquad
\end{eqnarray}

\noindent
(Note that in the second term the rightmost $\beta(\alpha)/\alpha$ coming from the series cancels $\alpha/\beta(\alpha)$ which is multiplied by $\gamma(\alpha)$, and we obtain the derivative $\partial/\partial\ln\alpha$ acting only on $\gamma(\alpha)$.) After calculating the sums over $p$ and $q$ the expression under consideration can be presented in the form

\begin{eqnarray}
&& \frac{\partial\ln Z}{\partial\ln\alpha} = - \exp\Big\{ \ln\frac{\Lambda}{\mu}\, \frac{\hat\partial}{\partial\ln\alpha} \frac{\beta(\alpha)}{\alpha}\Big\}\,
\bigg[\frac{\gamma(\alpha)}{\varepsilon}\Big(1-\frac{\beta(\alpha)}{\varepsilon\alpha}\Big)^{-1}\bigg] \nonumber\\
&& \qquad\qquad\qquad\qquad\qquad\qquad\qquad\quad
- \Big[\exp\Big\{ \ln\frac{\Lambda}{\mu}\, \frac{\hat\partial}{\partial\ln\alpha} \frac{\beta(\alpha)}{\alpha}\Big\} -1\Big] \frac{\alpha\gamma(\alpha)}{\beta(\alpha)}.\qquad
\end{eqnarray}

\noindent
Summing up the terms containing the exponential functions we obtain the final equation which encodes all higher $\varepsilon$-poles and logarithms in $\ln Z$,

\begin{equation}\label{LnZM_Final_Result}
\frac{\partial\ln Z}{\partial\ln\alpha} = \frac{\alpha\gamma(\alpha)}{\beta(\alpha)} - \exp\Big\{ \ln\frac{\Lambda}{\mu}\, \frac{\hat\partial}{\partial\ln\alpha} \frac{\beta(\alpha)}{\alpha}\Big\}\,
\bigg[\frac{\alpha\gamma(\alpha)}{\beta(\alpha)} \Big(1-\frac{\beta(\alpha)}{\varepsilon\alpha}\Big)^{-1}\bigg].
\end{equation}

\noindent
To check the correctness of this equation, we have again derived the five-loop expression (\ref{LnZM_5Loops}) for $\ln Z$ directly from Eq. (\ref{LnZM_Final_Result}). Also we note that after the formal replacement $\gamma(\alpha)\to \beta(\alpha)/\alpha$ Eq. (\ref{LnZM_Final_Result}) produces the corresponding expression for $\ln Z_\alpha$ given by Eq. (\ref{LnZ_Final_Result}).

Again, for completeness, we present the expressions for pure $\varepsilon$-poles and pure logarithms following from Eq. (\ref{LnZM_5Loops}). The result for pure poles is written as

\begin{eqnarray}\label{LnZM_5Loops_Poles}
&&\hspace*{-5mm} \ln Z\Big|_{\mu=\Lambda} = - \frac{\alpha \gamma_1}{\varepsilon} - \frac{\alpha^2}{2} \Big( \frac{\gamma_2}{\varepsilon}
+ \frac{\gamma_1\beta_1 }{\varepsilon^2}  \Big) - \frac{\alpha^3}{3} \Big( \frac{\gamma_3}{\varepsilon} + \frac{\gamma_1\beta_2 + \gamma_2\beta_1}{\varepsilon^2} + \frac{\gamma_1\beta_1^2}{\varepsilon^3} \Big)
\nonumber\\
&&\hspace*{-5mm} - \frac{\alpha^4}{4} \Big( \frac{\gamma_4}{\varepsilon} + \frac{\gamma_1\beta_3 +\gamma_2\beta_2 +\gamma_3\beta_1}{\varepsilon^2}
+ \frac{2 \gamma_1\beta_1\beta_2 + \gamma_2\beta_1^2}{\varepsilon^3} + \frac{\gamma_1\beta_1^3}{\varepsilon^4} \Big)
\nonumber\\
&&\hspace*{-5mm}
- \frac{\alpha^5}{5} \Big( \frac{\gamma_5}{\varepsilon} + \frac{\gamma_1\beta_4  + \gamma_2\beta_3 + \gamma_3\beta_2+\gamma_4\beta_1}{\varepsilon^2}
+ \frac{2\gamma_1\beta_1\beta_3 + \gamma_1\beta_2^2 + 2 \gamma_2\beta_1\beta_2 + \gamma_3\beta_1^2}{\varepsilon^3}
\nonumber\\
&&\hspace*{-5mm}
+ \frac{\gamma_2\beta_1^3 + 3\gamma_1\beta_1^2\beta_2}{\varepsilon^4} + \frac{\gamma_1\beta_1^4}{\varepsilon^5}
\Big) + O(\alpha^6).
\end{eqnarray}

\noindent
Up to notation, it agrees with the three-loop expression for $Z$ presented in \cite{Kastening:1996nj}. The corresponding result for pure logarithms is written as

\begin{eqnarray}\label{LnZM_5Loops_Logarithms}
&&\hspace*{-5mm} \ln Z\Big|_{\varepsilon^{-1}\to 0} = - \alpha \gamma_1 \ln\frac{\Lambda}{\mu} - \alpha^2 \Big( \gamma_2 \ln\frac{\Lambda}{\mu} + \frac{\gamma_1\beta_1}{2}\ln^2\frac{\Lambda}{\mu} \Big)
- \alpha^3 \Big(\gamma_3 \ln\frac{\Lambda}{\mu} + \frac{\gamma_1\beta_2 + 2\gamma_2\beta_1}{2}
\nonumber\\
&&\hspace*{-5mm} \times \ln^2\frac{\Lambda}{\mu} + \frac{\gamma_1\beta_1^2}{3} \ln^3\frac{\Lambda}{\mu}\Big) - \alpha^4 \Big( \gamma_4 \ln\frac{\Lambda}{\mu}
+ \frac{1}{2}\big(\gamma_1\beta_3 + 2 \gamma_2\beta_2 + 3\gamma_3\beta_1\big) \ln^2\frac{\Lambda}{\mu}
+ \frac{1}{6} \big( 5\gamma_1\beta_1\beta_2   \nonumber\\
&&\hspace*{-5mm} + 6\gamma_2\beta_1^2 \big) \ln^3\frac{\Lambda}{\mu} + \frac{\gamma_1\beta_1^3}{4} \ln^4\frac{\Lambda}{\mu} \Big)
- \alpha^5 \Big( \gamma_5 \ln \frac{\Lambda}{\mu}
+ \frac{1}{2} \big(\gamma_1\beta_4 + 2\gamma_2\beta_3 + 3\gamma_3\beta_2 + 4 \gamma_4\beta_1\big) \ln^2\frac{\Lambda}{\mu}
\nonumber\\
&&\hspace*{-5mm}  + \frac{1}{6}  \big(6\gamma_1\beta_1\beta_3 + 3\gamma_1\beta_2^2 +14 \gamma_2\beta_1\beta_2 + 12 \gamma_3\beta_1^2\big) \ln^3\frac{\Lambda}{\mu}
+ \frac{1}{12} \big(13\gamma_1\beta_1^2\beta_2 + 12 \gamma_2\beta_1^3\big) \ln^4\frac{\Lambda}{\mu}\nonumber\\
&&\hspace*{-5mm} + \frac{\gamma_1\beta_1^4}{5} \ln^5\frac{\Lambda}{\mu} \Big) + O(\alpha^6).
\end{eqnarray}

\noindent
As earlier, we should recall that $\beta_L$ and $\gamma_L$ are scheme dependent starting from $L\geq 3$ and $L\geq 2$, respectively.

\section{Relations between coefficients at $\varepsilon$-poles and logarithms}
\label{Section_Relations}

\subsection{How to transform $\varepsilon$-poles into logarithms}
\hspace*{\parindent}

The explicit expressions for $\ln Z_\alpha$, $(Z_\alpha)^S$, and $\ln Z$ derived above allow establishing the correspondence between the coefficients at $\varepsilon$-poles and logarithms. Namely, let us assume that we have expressed one of these values in the standard $\mbox{MS}$ (or $\overline{\mbox{MS}}$) scheme in terms of the coefficients $\beta_L$ and $\gamma_L$. Then it is possible to construct the corresponding expression in the HD+MSL scheme, when the renormalization constants contain only pure logarithms (certainly under the assumption that all divergences are logarithmic). Note that the coefficients of the $\beta$-function and of the anomalous dimension certainly depend on the renormalization scheme (starting from the three- and two-loop approximation, respectively). Therefore, in order to restore the HD+MSL result from the $\mbox{MS}$ result, one should take into account the change of the coefficients in Eqs. (\ref{Beta_Perturbative_Expansion}) and (\ref{Gamma_Perturbative_Expansion}). However, here we will only investigate how the dependence of the renormalization constants on $\beta_L$ and $\gamma_L$ changes if we transform $\varepsilon$-poles into logarithms (for the $\mbox{MS}$-like renormalization prescriptions).

Let us start with the expression (\ref{lnZ_Final_Expansion}) for $\ln Z_\alpha$. Using Eq. (\ref{Z_Alpha_Equality}) we see that in the standard $\mbox{MS}$ scheme (without logarithms)

\begin{equation}\label{LnZM_MS}
\ln \bm{Z_\alpha}(\alpha,1/\varepsilon) = \ln Z_{\alpha}(\alpha,1/\varepsilon,\ln\Lambda/\mu)\Big|_{\mu=\Lambda} = - \sum\limits_{q=1}^\infty\, \sum\limits_{k_1,k_2,\ldots, k_{q}=1}^\infty \frac{1}{K_{q}}\, \beta_{k_1} \beta_{k_2} \ldots \beta_{k_{q}}\, \alpha^{K_{q}}\, \varepsilon^{-q}.
\end{equation}

\noindent
From the other side, in the HD+MSL scheme (in which only pure logarithms are present in the renormalization constants) the analogous equation takes the form

\begin{equation}
\ln Z_{\alpha}(\alpha,1/\varepsilon,\ln\Lambda/\mu)\Big|_{\varepsilon^{-1}\to 0} = - \sum\limits_{p=1}^\infty \sum\limits_{k_1,k_2,\ldots, k_{p}=1}^\infty
\frac{1}{K_{p}}\cdot \frac{K_{p}\bm{!}}{p!}\, \beta_{k_1} \beta_{k_2} \ldots \beta_{k_{p}}\, \alpha^{K_{p}}\, \ln^p \frac{\Lambda}{\mu}.
\end{equation}

\noindent
Making in this equation the replacement $p\to q$ and comparing it with Eq. (\ref{LnZM_MS}) we see that the HD+MSL result can be obtained from the $\mbox{MS}$ one after the replacement $\varepsilon^{-1} \to \ln\Lambda/\mu$ by inserting the factor $K_q\bm{!}/q!$.

The expressions for $(Z_\alpha)^S$ and $\ln Z$ are considered similarly. The result is exactly the same. Thus, for $X=\{\ln Z_\alpha,\, (Z_\alpha)^S,\, \ln Z \}$ if in the $\mbox{MS}$ scheme a certain expression is given by the series

\begin{equation}
X\Big|_{\text{MS}} = \sum_{n=1}^\infty \sum\limits_{k_1,\ldots, k_n=1}^\infty X_{k_1\ldots k_n} \alpha^{K_n} \varepsilon^{-n},
\end{equation}

\noindent
then in the HD+MSL scheme the corresponding series takes the form

\begin{equation}
X\Big|_{\text{HD+MSL}} = \sum_{n=1}^\infty \sum\limits_{k_1,\ldots, k_n=1}^\infty \frac{K_n\bm{!}}{n!} X_{k_1\ldots k_n} \alpha^{K_n} \ln^n\frac{\Lambda}{\mu},
\end{equation}

\noindent
where $K_n\bm{!}$ is defined by Eq. (\ref{Generalized_Factorial}). Certainly, it is also necessary to take into account that the coefficients of the $\beta$-function and anomalous dimension in the HD+MSL and $\mbox{MS}$ schemes are different.

\subsection{Some features of $\ln Z_\alpha$}
\hspace*{\parindent}\label{Subsection_LnZ_Features}

From the explicit five-loop expression for $\ln Z_\alpha$ given by Eq. (\ref{LnZ_5Loops}) in Appendix \ref{Appendix_Explicit_Z} we see that in this order all terms proportional to $1/\varepsilon^2$, $\varepsilon^{-1} \ln\Lambda/\mu$, and $\ln^2\Lambda/\mu$ are factorized into perfect squares. Here starting from the general equation (\ref{lnZ_Final_Expansion}) we demonstrate that this feature is valid in all orders of the perturbation theory. According to Eq. (\ref{LnZ_Result}), the coefficient at $1/\varepsilon^2$ in $L$ loops is

\begin{equation}
\widetilde B_{L,\,0,\,2} = -\frac{1}{L} \sum\limits_{k_1+k_2=L} \beta_{k_1} \beta_{k_2} = -\frac{1}{L}\sum\limits_{k=1}^{L-1} \beta_k \beta_{L-k}.
\end{equation}

\noindent
Similarly, the coefficient at $\varepsilon^{-1} \ln\Lambda/\mu$ is written as

\begin{equation}
\widetilde B_{L,\,1,\,1} = -\frac{1}{L} \sum\limits_{k_1+k_2=L} \beta_{k_1} \beta_{k_2} (k_1 + k_2) = - \sum\limits_{k=1}^{L-1} \beta_k \beta_{L-k},
\end{equation}

\noindent
and the coefficient at $\ln^2\Lambda/\mu$ has the form

\begin{eqnarray}
&& \widetilde B_{L,\,2,\,0} = -\frac{1}{2L} \sum\limits_{k_1+k_2=L} \beta_{k_1} \beta_{k_2} (k_1+k_2) k_1 = -\frac{1}{2} \sum\limits_{k_1+k_2=L} \beta_{k_1} \beta_{k_2} k_1 \nonumber\\
&&\qquad\qquad\qquad\qquad\qquad\qquad
=  -\frac{1}{4} \sum\limits_{k_1+k_2=L} \beta_{k_1} \beta_{k_2} (k_1+k_2) = -\frac{L}{4} \sum\limits_{k=1}^{L-1} \beta_k \beta_{L-k}.\qquad
\end{eqnarray}

\noindent
Therefore, the terms under consideration give the perfect square

\begin{equation}
-\frac{1}{L}\sum\limits_{k=1}^{L-1} \beta_k \beta_{L-k} \Big(\frac{1}{\varepsilon^2} + \frac{L}{\varepsilon} \ln\frac{\Lambda}{\mu} + \frac{L^2}{4}\ln^2\frac{\Lambda}{\mu}\Big) = - \frac{1}{L}\sum\limits_{k=1}^{L-1} \beta_k \beta_{L-k}\Big(\frac{1}{\varepsilon} + \frac{L}{2}\ln\frac{\Lambda}{\mu}\Big)^2.
\end{equation}

\noindent
This implies that for $\ln Z_\alpha$ it is possible to write down the expression generalizing Eq. (\ref{LnZ_Lowest_Structure})

\begin{eqnarray}
&& \ln Z_\alpha = -\sum\limits_{L=1}^\infty \alpha^L \beta_L \Big(\frac{1}{L\varepsilon} + \ln\frac{\Lambda}{\mu}\Big) - \frac{1}{L}\sum\limits_{L=2}^\infty \alpha^L \sum\limits_{k=1}^{L-1} \beta_k \beta_{L-k}\Big(\frac{1}{\varepsilon} + \frac{L}{2}\ln\frac{\Lambda}{\mu}\Big)^2\qquad
\nonumber\\
&&\qquad\qquad\qquad\qquad\qquad\qquad\qquad\qquad\qquad\qquad +\ \mbox{higher poles and logarithms},\qquad
\end{eqnarray}

\noindent
which perfectly agrees with Eq. (\ref{LnZ_5Loops}).

Note that a similar structure appears in the terms with coinciding $k_i$. Really, if $k_1=\ldots =k_{p+q}\equiv k$, then from Eq. (\ref{lnZ_Final_Expansion}) we see that $L=(p+q)k$ and the corresponding contribution to $\ln Z_\alpha$ takes the form

\begin{equation}
\ln Z_\alpha = - \sum\limits_{\stackrel{\scriptsize \mbox{$p,q=0$}}{p+q\geq 1}}^\infty\, \sum\limits_{k=1}^\infty \frac{(p+q-1)!}{p!\, q!} k^{p-1} \beta_k^{p+q} \alpha^{(p+q)k} \varepsilon^{-q} \ln^p\frac{\Lambda}{\mu} + \mbox{the other terms}.
\end{equation}

\noindent
Introducing the new summation index $m\equiv p+q$ and using the binomial theorem this expression can be presented in the form

\begin{eqnarray}\label{Z_Alpha_Expression}
&&\hspace*{-3mm} \ln Z_\alpha = - \sum\limits_{m=1}^\infty \sum\limits_{k=1}^\infty \frac{1}{mk}\, \sum\limits_{p=0}^m C^p_m k^p \beta_k^m \alpha^{mk} \varepsilon^{-m+p} \ln^p\frac{\Lambda}{\mu} + \mbox{the other terms}\qquad \nonumber\\
&&\hspace*{-3mm}\qquad\qquad\qquad\qquad\qquad = - \sum\limits_{m=1}^\infty \sum\limits_{k=1}^\infty \frac{\big(\beta_k \alpha^{k}\big)^m}{mk} \Big(\frac{1}{\varepsilon} + k\ln\frac{\Lambda}{\mu}\Big)^m + \mbox{the other terms},\qquad
\end{eqnarray}

\noindent
where

\begin{equation}
C^p_m\equiv \frac{m!}{p!(m-p)!}
\end{equation}

\noindent
are the binomial coefficients. Again, it is easy to verify that the terms included into Eq. (\ref{Z_Alpha_Expression}) exactly agree with the explicit five-loop result (\ref{LnZ_5Loops}).

\subsection{Some features of $Z_\alpha$ and $(Z_\alpha)^S$}
\hspace*{\parindent}

Looking at the explicit six-loop expression (\ref{Z_Alpha_6Loops}) for the renormalization constant $Z_\alpha$ presented in Appendix \ref{Appendix_Explicit_Z} we see that all terms containing the only $\beta_L$ with $L\geq 2$ (in Eq. (\ref{Z_Alpha_6Loops}) they as well as the one-loop contribution are marked by the bold font) are factorized into the structures

\begin{equation}\label{Interesting_Structure}
\varepsilon^{L-k} \Big(\frac{1}{\varepsilon} + \ln\frac{\Lambda}{\mu} \Big)^L\bigg|_{\varepsilon^s\to 0\ \text{for all}\ s>0} = \sum\limits_{p=0}^k C_L^p\, \varepsilon^{-k+p} \ln^{p}\frac{\Lambda}{\mu}
= \sum\limits_{p=0}^k \frac{L!}{p! (L-p)!}\, \varepsilon^{-k+p} \ln^{p}\frac{\Lambda}{\mu},
\end{equation}

\noindent
where $L$ is a number of loops and $1 \leq k \leq L-1$. As earlier, the condition ``$\varepsilon^s\to 0\ \text{for all}\ s>0$'' implies that all terms proportional to the positive powers of $\varepsilon$ should be excluded from the considered expression.

The statement formulated above can be proven in all orders of the perturbation theory. As a starting point of the proof we consider the expression (\ref{ZS_Final_Expansion}) with $S=1$,

\begin{equation}\label{Z_Alpha_Expansion}
Z_\alpha = 1 - \sum\limits_{\stackrel{\scriptsize \mbox{$p,q=0$}}{p+q\geq 1}}^\infty\, \sum\limits_{k_1,k_2,\ldots, k_{p+q}=1}^\infty \frac{(-1+K_{p+q-1})\bm{!}}{p!\, K_q\bm{!}}\, \beta_{k_1} \beta_{k_2} \ldots \beta_{k_{p+q}}\, \alpha^{K_{p+q}}\, \varepsilon^{-q}\, \ln^p \frac{\Lambda}{\mu}.
\end{equation}

\noindent
According to Eq. (\ref{Generalized_Factorial_With_S}),

\begin{equation}\label{S=1_Factorial}
\quad(-1+K_{p+q-1})\bm{!} = (-1+k_1) (-1+k_1+k_2)\times \ldots\times (-1+k_1+k_2+\ldots + k_{p+q-1}),\quad
\end{equation}

\noindent
so that the terms in which all $k_i=1$ give vanishing contributions except for the ones corresponding to $p+q=1$, which give the contribution

\begin{equation}\label{Z_Alpha_One_loop}
- \alpha \beta_1 \Big(\frac{1}{\varepsilon} + \ln\frac{\Lambda}{\mu}\Big).
\end{equation}

\noindent
Next, let us consider the terms in that all $k_i$ are equal to 1 except for one, which is equal to $m>1$. From Eq. (\ref{S=1_Factorial}) it is evident that a nontrivial contribution to $Z_\alpha$ appears only if $k_1$ differs from 1, while the others $k_i$ (with $i\geq 2$) are equal to 1. Taking into account that $k_1+k_2+\ldots +k_{p+q} = L\geq 2$ we conclude that $k_1=m = L-p-q+1$. In this case

\begin{eqnarray}
&& \frac{(-1+K_{p+q})\bm{!}}{p!\, K_q\bm{!}\, (-1+K_{p+q})} = \frac{(m-1)m\ldots (m+q+p-3)}{p!\, m\ldots (m+q-1)}\nonumber\\
&&\qquad\qquad\qquad\qquad = (m-1)\frac{(m+q+p-3)!}{p!\,(m+q-1)!} = \frac{L-q-p}{L(L-1)} \cdot \frac{L!}{p! (L-p)!}.\qquad
\end{eqnarray}

\noindent
Substituting these values into Eq. (\ref{Z_Alpha_Expansion}) and including the one-loop contribution (\ref{Z_Alpha_One_loop}) we obtain the expression for the renormalization constant $Z_\alpha$ in the form

\begin{eqnarray}\label{Z_Alpha_Second_Sequence}
&& Z_\alpha = 1 - \alpha \beta_1 \Big(\frac{1}{\varepsilon} + \ln\frac{\Lambda}{\mu}\Big) - \sum\limits_{\stackrel{\scriptsize \mbox{$p,q=0$}}{p+q\geq 1}}^\infty\, \sum\limits_{L=1+p+q}^\infty \frac{L-q-p}{L(L-1)} \cdot \frac{L!}{p! (L-p)!} (\beta_1)^{p+q-1} \beta_{L-q-p+1} \qquad\nonumber\\
&& \times \alpha^{L}\, \varepsilon^{-q}\, \ln^p \frac{\Lambda}{\mu} + \mbox{terms in which at least two $k_i > 1$}.
\end{eqnarray}

\noindent
(Note that the first term in which two $k_i\ne 1$ is proportional to $(\beta_2)^2$ and appears only in the four-loop approximation, see the explicit six-loop expression (\ref{Z_Alpha_6Loops}).) Introducing the new summation index $k\equiv p+q$ and taking into account that $1\leq p+q\leq L-1$ we rewrite Eq. (\ref{Z_Alpha_Second_Sequence}) in the form

\begin{eqnarray}
&&\hspace*{-5mm} Z_\alpha = 1 - \alpha \beta_1 \Big(\frac{1}{\varepsilon} + \ln\frac{\Lambda}{\mu}\Big) - \sum\limits_{L=2}^\infty \alpha^{L}\, \sum\limits_{k=1}^{L-1}\, \frac{L-k}{L(L-1)} (\beta_1)^{k-1} \beta_{L-k+1}
\nonumber\\
&&\hspace*{-5mm}\qquad\qquad
\times \sum\limits_{p=0}^k \frac{L!}{p! (L-p)!}
\varepsilon^{-k+p}\, \ln^p \frac{\Lambda}{\mu}\ +\ \mbox{terms in which at least two $k_i\ne 1$}.\qquad
\end{eqnarray}

\noindent
With the help of the binomial theorem this expression can be transformed to the final form

\begin{eqnarray}\label{Z_Alpha_Main_Sequence}
&&\hspace*{-5mm} Z_\alpha = 1 - \alpha \beta_1 \Big(\frac{1}{\varepsilon} + \ln\frac{\Lambda}{\mu}\Big) - \sum\limits_{L=2}^\infty \frac{\alpha^{L}}{L(L-1)} \sum\limits_{k=1}^{L-1}\, (L-k) (\beta_1)^{k-1} \beta_{L-k+1}\, \qquad\nonumber\\
&&\hspace*{-5mm} \qquad\quad \times \varepsilon^{L-k}\, \Big(\frac{1}{\varepsilon} + \ln\frac{\Lambda}{\mu}\Big)^L\bigg|_{\varepsilon^s\to 0\ \text{for all}\ s>0} + \mbox{terms in which at least two $k_i\ne 1$}.\qquad
\end{eqnarray}

\noindent
In the sixth order this equation gives the expansion

\begin{eqnarray}\label{Z_Alpha_Main_Sequence_6Loops}
&&\hspace*{-5mm} Z_\alpha = \bigg[1 - \alpha \beta_1 \Big(\frac{1}{\varepsilon} + \ln\frac{\Lambda}{\mu}\Big) - \frac{\alpha^2}{2} \varepsilon \beta_2 \Big(\frac{1}{\varepsilon} + \ln\frac{\Lambda}{\mu}\Big)^2
- \frac{\alpha^3}{6} \Big(2\varepsilon^2\beta_3 + \varepsilon \beta_1\beta_2\Big)\Big(\frac{1}{\varepsilon} + \ln\frac{\Lambda}{\mu}\Big)^3 \nonumber\\
&&\hspace*{-5mm} - \frac{\alpha^4}{12} \Big(3\varepsilon^3\beta_4 + 2\varepsilon^2 \beta_1 \beta_3 + \varepsilon\beta_1^2 \beta_2\Big)\Big(\frac{1}{\varepsilon} + \ln\frac{\Lambda}{\mu}\Big)^4
\nonumber\\
&&\hspace*{-5mm} - \frac{\alpha^5}{20} \Big(4\varepsilon^4\beta_5 + 3\varepsilon^3 \beta_1 \beta_4 + 2\varepsilon^2 \beta_1^2 \beta_3 +\varepsilon \beta_1^3\beta_2\Big)\Big(\frac{1}{\varepsilon} + \ln\frac{\Lambda}{\mu}\Big)^5
\nonumber\\
&&\hspace*{-5mm} - \frac{\alpha^6}{30} \Big(5\varepsilon^5\beta_6 + 4\varepsilon^4 \beta_1 \beta_5 + 3\varepsilon^3\beta_1^2 \beta_4 + 2\varepsilon^2\beta_1^3\beta_3 + \varepsilon\beta_1^4\beta_2\Big)\Big(\frac{1}{\varepsilon} + \ln\frac{\Lambda}{\mu}\Big)^6
\bigg]\bigg|_{\varepsilon^s\to 0\ \text{for all}\ s>0}\nonumber\\
&&\hspace*{-5mm} +\ \mbox{terms in which at least two $k_i\ne 1$} + O(\alpha^7),\vphantom{\Big(}
\end{eqnarray}

\noindent
which exactly reproduces the terms that are indicated by the bold font in Eq. (\ref{Z_Alpha_6Loops}).

For an arbitrary $S$ it is also possible to use the binomial theorem for the terms in which all $k_i$ coincide. Starting from Eq. (\ref{ZS_Final_Expansion}) and repeating the argumentation of Section \ref{Subsection_LnZ_Features} after some simple transformations we obtain that the terms of the considered structure are given by the series

\begin{eqnarray}\label{ZS_Second_Sequence}
&& (Z_\alpha)^S = 1 - S \sum\limits_{k=1}^\infty \frac{\beta_k \alpha^{k}}{k} \Big(\frac{1}{\varepsilon}+k\ln\frac{\Lambda}{\mu}\Big)
- S \sum\limits_{k=1}^\infty \sum\limits_{m=2}^\infty \frac{\big(\beta_k \alpha^{k}\big)^m}{k^m\,m!} (-S+k)(-S+2k) \qquad\nonumber\\
&& \times \ldots \times (-S+(m-1)k)\, \Big(\frac{1}{\varepsilon}+k\ln\frac{\Lambda}{\mu}\Big)^m + \mbox{the other terms}.\qquad
\end{eqnarray}

\noindent
Comparing it with Eqs. (\ref{ZS_5Loops}) and (\ref{Z_Alpha_6Loops}) we see that this expansion exactly agrees with the explicit expressions.

Now, let us consider the terms with the highest overall degree of $1/\varepsilon$ and $\ln\Lambda/\mu$ in an $L$-loop approximation assuming that $L\geq 2$. (This degree is certainly equal to $p+q$.)  For a fixed $K_{p+q}=k_1+k_2+\ldots+k_{p+q} =L$ the maximal value of $p+q$ corresponds to the case when $k_i$ take the minimal possible values. For $Z_\alpha$ this implies that one $k_i$ is equal to 2 and the others are equal to 1, so that $p+q=L-1$. Therefore, the terms with the highest poles and logarithms are proportional to $\beta_2 \beta_1^{L-2}$. It is well known \cite{Vladimirov:1979ib,Vladimirov:1979my} that the coefficients $\beta_1$ and $\beta_2$ are scheme independent, so that the terms with the largest ($L-1$) overall degree of $1/\varepsilon$ and $\ln\Lambda/\mu$ contain only scheme independent coefficients of the $\beta$-function.

For $\ln Z_\alpha$ and $(Z_\alpha)^S$ (where $S\geq L$ or $S$ is not a positive integer) there are also terms of degree $L$, in which the coefficients are proportional to $\beta_1^L$. Therefore, in this case both leading and (the first) subleading terms contain only scheme independent coefficients of the $\beta$-function. However, for the positive integer $S<L$ from Eq. (\ref{ZS_Second_Sequence}) we see that the terms with the degree $L$ (corresponding to $k=1$, $m=L$) disappear. In this case only leading terms of degree $(L-1)$ contain only scheme independent coefficients of the $\beta$-function exactly as for $Z_\alpha$.

\subsection{Some features of $\ln Z$}
\hspace*{\parindent}

Let us now investigate the features of $\ln Z$. From the five-loop expression (\ref{LnZM_5Loops}) we see that the coefficients at the terms which contain only powers of $\beta_1$ (and do not contain $\beta_k$ with $k\geq 2$) have the form (\ref{Interesting_Structure}). Exactly as earlier, it is possible to prove this fact in all loops and find explicit expressions for these coefficients. As a starting point we consider the exact expression (\ref{lnZM_Final_Expansion}) and look at the terms containing only powers of $\beta_1$. They are obtained if $k_i=1$ for all $i\geq 2$ and $k_1 = m\geq 1$. Taking into account that $k_1+k_2+\ldots + k_{q+p} = L\geq 1$, where $L$ is a number of loops, we see that $m = L-q-p+1$. Then the coefficient at the term under consideration takes the form

\begin{eqnarray}
&& \frac{K_{p+q}\bm{!}}{K_{p+q}\, p!\, K_q\bm{!}} = \frac{m(m+1)\ldots (m+q+p-1)}{L\,p!\, m\ldots (m+q-1)}\nonumber\\
&&\qquad\qquad\qquad\qquad\qquad\qquad = \frac{(m+q+p-1)!}{L\,p!\,(m+q-1)!} = \frac{1}{L} \cdot \frac{L!}{p! (L-p)!}.\qquad
\end{eqnarray}

\noindent
Substituting this expression into Eq. (\ref{lnZM_Final_Expansion}) we can present $\ln Z$ as

\begin{eqnarray}\label{LnZM_Auxiliary}
&& \ln Z =  - \smash{\sum\limits_{\stackrel{\scriptsize \mbox{$p,q=0$}}{p+q\geq 1}}^\infty}\, \sum\limits_{L=p+q}^\infty \frac{\alpha^L}{L}\, \frac{L!}{p! (L-p)!} (\beta_1)^{p+q-1} \gamma_{L-p-q+1} \varepsilon^{-q}\, \ln^p \frac{\Lambda}{\mu}
\qquad\nonumber\\
&&\qquad\qquad\qquad\qquad\qquad\qquad\qquad\qquad\qquad +\ \mbox{terms containing $\beta_i$ with $i\geq 2$}.\vphantom{\Big(}\qquad
\end{eqnarray}

\noindent
Introducing $k\equiv p+q$ and taking into account that in this case $1 \leq p+q\leq L$ we rewrite Eq. (\ref{LnZM_Auxiliary}) in the form

\begin{eqnarray}
&& \ln Z =  - \sum\limits_{L=1}^\infty \frac{\alpha^L}{L}\, \sum\limits_{k=1}^{L}\, \gamma_{L-k+1} (\beta_1)^{k-1} \sum\limits_{p=0}^k \frac{L!}{p! (L-p)!}
\varepsilon^{-k+p}\, \ln^p \frac{\Lambda}{\mu}\nonumber\\
&&\qquad\qquad\qquad\qquad\qquad\qquad\qquad\qquad\qquad +\ \mbox{terms containing $\beta_i$ with $i\geq 2$}.\vphantom{\Big(}\qquad
\end{eqnarray}

\noindent
Next, using the binomial theorem we obtain the required structure

\begin{eqnarray}\label{LnZM_Main_Sequence}
&& \ln Z = - \sum\limits_{L=1}^\infty \frac{\alpha^{L}}{L}\, \sum\limits_{k=1}^{L}\, \gamma_{L-k+1} (\beta_1)^{k-1} \varepsilon^{L-k}\, \Big(\frac{1}{\varepsilon} + \ln\frac{\Lambda}{\mu}\Big)^L\bigg|_{\varepsilon^s\to 0\ \text{for all}\ s>0} \qquad\nonumber\\
&&\qquad\qquad\qquad\qquad\qquad\qquad\qquad\qquad\qquad +\ \mbox{terms containing $\beta_i$ with $i\geq 2$}.\qquad
\end{eqnarray}

\noindent
In the first five loops this equation gives the expansion

\begin{eqnarray}\label{LnZM_Main_Sequence_5Loops}
&&\hspace*{-3mm} \ln Z = \bigg[ - \alpha \gamma_1 \Big(\frac{1}{\varepsilon} + \ln\frac{\Lambda}{\mu}\Big) - \frac{\alpha^2}{2} \Big(\varepsilon \gamma_2 +\gamma_1 \beta_1 \Big)\Big(\frac{1}{\varepsilon} + \ln\frac{\Lambda}{\mu}\Big)^2
- \frac{\alpha^3}{3} \Big(\varepsilon^2\gamma_3 + \varepsilon \gamma_2\beta_1 + \gamma_1\beta_1^2\Big) \nonumber\\
&&\hspace*{-3mm} \times \Big(\frac{1}{\varepsilon} + \ln\frac{\Lambda}{\mu}\Big)^3 - \frac{\alpha^4}{4} \Big(\varepsilon^3 \gamma_4 + \varepsilon^2 \gamma_3 \beta_1 + \varepsilon \gamma_2 \beta_1^2 + \gamma_1 \beta_1^3\Big)\Big(\frac{1}{\varepsilon} + \ln\frac{\Lambda}{\mu}\Big)^4 - \frac{\alpha^5}{5} \Big(\varepsilon^4 \gamma_5 + \varepsilon^3 \gamma_4 \beta_1
\nonumber\\
&&\hspace*{-3mm} + \varepsilon^2 \gamma_3 \beta_1^2 + \varepsilon \gamma_2 \beta_1^3 +\gamma_1 \beta_1^4\Big)\Big(\frac{1}{\varepsilon} + \ln\frac{\Lambda}{\mu}\Big)^5
\bigg]\bigg|_{\varepsilon^s\to 0\ \text{for all}\ s>0}\nonumber\\
&&\hspace*{-3mm} \qquad\qquad\qquad\qquad\qquad\qquad\quad +\ \mbox{terms containing $\beta_i$ with $i\geq 2$} + O(\alpha^6),\vphantom{\Big(}
\end{eqnarray}

\noindent
which perfectly agrees with Eq. (\ref{LnZM_5Loops}).

Again we note that the terms in $\ln Z$ with the highest overall degree of $1/\varepsilon$ and $\ln\Lambda/\mu$ (in a given order of the perturbation theory) contain only scheme independent coefficients of RGFs. Really, in $L$ loops they have the degree $L$ and are proportional to $\gamma_1\beta_1^{L-1}$. Taking into account that the one-loop contributions to the $\beta$-function and to the anomalous dimension are scheme independent we obtain the required statement.

Note that it is also possible to find the sum of all terms with coinciding $k_i$. Exactly as in Section \ref{Subsection_LnZ_Features} we obtain the series

\begin{equation}
\ln Z = - \sum\limits_{m=1}^\infty \sum\limits_{k=1}^\infty \frac{\gamma_k \beta_k^{m-1} \alpha^{mk}}{mk} \Big(\frac{1}{\varepsilon} + k\ln\frac{\Lambda}{\mu}\Big)^m + \mbox{the other terms},
\end{equation}

\noindent
which exactly agrees with the explicit five-loop expression (\ref{LnZM_5Loops}).

\section{Examples}
\hspace*{\parindent}\label{Section_Examples}

In this section we compare the general expressions presented in the previous sections with the results of some explicit calculations made earlier.

\subsection{${\cal N}=1$ supersymmetric quantum electrodynamics}
\hspace*{\parindent}\label{Subsection_N=1_SQED}

First we consider ${\cal N}=1$ SQED with $N_f$ flavors, which in the massless limit is described by the (superfield) action

\begin{equation}
S = \frac{1}{4e_0^2}\mbox{Re} \int d^4x\,d^2\theta\, W^a W_a + \sum\limits_{\alpha=1}^{N_f} \frac{1}{4} \int d^4x\, d^4\theta\,\Big(\phi_\alpha^* e^{2V} \phi_\alpha + \widetilde\phi_\alpha^* e^{-2V}\widetilde\phi_\alpha \Big),
\end{equation}

\noindent
where $V$ is the gauge superfield, $\phi_\alpha$ and $\widetilde\phi_\alpha$ are $N_f$ pairs of the chiral matter superfields, and $W_a = \bar D^2 D_a V/4$ is a supersymmetric gauge superfield strength. The bare gauge coupling constant is defined as $\alpha_0=e_0^2/4\pi$, and the matter superfields (for all $\alpha=1,\ldots,N_f$) are renormalized as $\phi_{\alpha} = \sqrt{Z}\phi_{\alpha,R}$ and $\widetilde\phi_{\alpha}=\sqrt{Z}\widetilde\phi_{\alpha,R}$, where the subscript $R$ indicates the renormalized superfields.

Using the version of the dimensional technique considered in this paper (analogous to the $\overline{\mbox{DR}}$ scheme) the three-loop $Z_\alpha$ and the two-loop $\ln Z$ have been calculated in \cite{Aleshin:2016rrr}.\footnote{The contribution to $Z_\alpha$ proportional to $\alpha^3 N_f$ was not found in \cite{Aleshin:2016rrr}. However, it can be obtained from the result of \cite{Aleshin:2019yqj} for the Adler $D$-function if one makes a certain formal replacement of the group Casimirs.} The results are given by the expressions

\begin{eqnarray}\label{N=1_SQED_Z_Alpha}
&& Z_\alpha = 1 - \frac{\alpha N_f}{\pi} \Big(\frac{1}{\varepsilon} +  \ln \frac{\bar\Lambda}{\mu} \Big) - \frac{\alpha^2 N_f}{\pi^2} \Big(\frac{1}{2 \varepsilon} +  \ln \frac{\bar\Lambda}{\mu} \Big) +\frac{\alpha^3 N_f}{\pi^3}\bigg[ \frac{1}{6\varepsilon} + \frac{1}{2}\ln\frac{\bar\Lambda}{\mu}
\nonumber\\	
&&\qquad\qquad\qquad\qquad\quad  - N_f \Big(- \frac{1}{4 \varepsilon} - \frac{3}{4} \ln \frac{\bar\Lambda}{\mu}
+ \frac{1}{6\varepsilon^2} + \frac{1}{2 \varepsilon}\ln\frac{\bar\Lambda}{\mu} + \frac{1}{2} \ln^2 \frac{\bar\Lambda}{\mu} \Big)\bigg] + O(\alpha^4);\qquad\nonumber\\
\label{N=1_SQED_Z}
&& \ln Z = \frac{\alpha}{\pi} \Big(\frac{1}{\varepsilon} +  \ln \frac{\bar\Lambda}{\mu} \Big)
+ \frac{\alpha^2}{\pi^2} \bigg[ -\frac{1}{4 \varepsilon} - \frac{1}{2} \ln\frac{\bar\Lambda}{\mu}
\nonumber\\
&&\qquad\qquad\qquad\qquad\quad\ \, + N_f \Big(-\frac{1}{4 \varepsilon} - \frac{1}{2}  \ln \frac{\bar\Lambda}{\mu} + \frac{1}{2 \varepsilon^2} + \frac{1}{\varepsilon} \ln \frac{\bar\Lambda}{\mu} + \frac{1}{2} \ln^2 \frac{\bar\Lambda}{\mu} \Big)\bigg] + O(\alpha^3).\qquad
\end{eqnarray}

\noindent
We see that the result for $Z_\alpha$ is in exact agreement with Eq. (\ref{Z_Alpha_6Loops}) (certainly, after the replacement $\Lambda\to \bar\Lambda$). Moreover, it satisfies Eq. (\ref{Z_Alpha_Main_Sequence_6Loops}), while the terms in which at least two $k_i$ are not equal to 1 do not appear in the considered approximation. Therefore, all relations which should be valid for higher poles and logarithms are satisfied. Comparing Eqs. (\ref{Z_Alpha_Main_Sequence_6Loops}) and (\ref{N=1_SQED_Z_Alpha}) we see that

\begin{equation}
\beta_1 = \frac{N_f}{\pi};\qquad \beta_2 = \frac{N_f}{\pi^2};\qquad \beta_3 = - \frac{2N_f +3N_f^2}{4 \pi^3}.
\end{equation}

\noindent
in agreement with \cite{Jack:1996vg,Jack:1996cn}, see also \cite{Kataev:2013csa,Kataev:2014gxa}.

Similarly, the expression for $\ln Z$ agrees with Eq. (\ref{LnZM_5Loops}) in which

\begin{equation}
\gamma_1 = -\frac{1}{\pi};\qquad \gamma_2 = \frac{1}{2 \pi^2} \big(1+  N_f\big).
\end{equation}

\noindent
Thus, all equations relating the coefficients at higher poles and logarithms to the coefficients in the perturbative expansions of the $\beta$-function and the anomalous dimension are satisfied in this case.

Also we see that the expression for $\ln Z_\alpha$ calculated on the base of Eq. (\ref{N=1_SQED_Z_Alpha})

\begin{eqnarray}
&&\hspace*{-5mm} \ln Z_\alpha = - \frac{\alpha N_f}{\pi} \Big(\frac{1}{\varepsilon} +  \ln \frac{\bar\Lambda}{\mu} \Big) - \frac{\alpha^2 N_f}{\pi^2} \bigg[\frac{1}{2 \varepsilon} +  \ln\frac{\bar\Lambda}{\mu}
+ \frac{N_f}{2} \Big( \frac{1}{\varepsilon} + \ln \frac{\bar\Lambda}{\mu}\Big)^2 \bigg] - \frac{\alpha^3 N_f}{\pi^3} \bigg[ -\frac{1}{6\varepsilon} \qquad\nonumber\\
&&\hspace*{-5mm}  - \frac{1}{2}\ln\frac{\bar\Lambda}{\mu} + N_f \bigg(- \frac{1}{4 \varepsilon} - \frac{3}{4} \ln \frac{\bar\Lambda}{\mu} + \frac{2}{3} \Big( \frac{1}{\varepsilon} + \frac{3}{2}\ln\frac{\bar\Lambda}{\mu}\Big)^2\bigg)
+ \frac{N_f^2}{3} \Big( \frac{1}{\varepsilon} + \ln\frac{\bar\Lambda}{\mu}\Big)^3 \bigg] + O(\alpha^4)\qquad
\end{eqnarray}

\noindent
agrees with Eq. (\ref{LnZ_5Loops}). This certainly confirms the correctness of the general equations derived above.

Note that for pure logarithms the agreement of the expressions (\ref{Z_Alpha_6Loops_Logarithms}) and (\ref{LnZM_5Loops_Logarithms}) with the result of the explicit four- and three-loop calculations made in \cite{Shirokov:2022jyd} has already been demonstrated in \cite{Meshcheriakov:2022tyi}.

\subsection{Coefficients at $\varepsilon$-poles in the $\varphi^4$-theory}
\hspace*{\parindent}\label{Subsection_Phi4}

The above results can also be verified by comparing them with explicit expressions for the renormalization constants of the $O(N)$-invariant $\varphi^4$-theory described by the Lagrangian

\begin{equation}\label{Varphi4_Lagrangian}
{\cal L} = \frac{1}{2} \sum\limits_{a=1}^N \Big(\partial_\mu\varphi_a \partial^\mu\varphi_a - m_0^2 \varphi_a^2\Big) - \frac{\lambda_0}{4!} \Big(\sum\limits_{a=1}^N \varphi_a^2 \Big)^2.
\end{equation}

\noindent
It is also convenient to introduce the new bare coupling constant

\begin{equation}
g_0\equiv \frac{\lambda_0}{16\pi^2}.
\end{equation}

\noindent
The renormalization constants for this theory are defined by the equations\footnote{Note that we use the notations in which the bare coupling and the bare field have integer dimensions.}

\begin{eqnarray}\label{Varphi4_Renormalization_Constants}
g_0 = g (Z_g)^{-1};\qquad \varphi =\sqrt{Z_\varphi} \varphi_R;\qquad  m_0 = \sqrt{Z_m} m,
\end{eqnarray}

\noindent
where the subscript $0$ indicates the bare coupling constant and the bare mass, while the subscript $R$ denotes the renormalized scalar field. The explicit five-loop expressions for these renormalization constants in a certain $\mbox{MS}$-like scheme can be found in \cite{Kleinert:2001ax}. It is similar to the $\overline{\mbox{MS}}$ scheme, but the substitution analogous to Eq. (\ref{MS_Bar_Mu}) is

\begin{equation}
\mu \to \frac{\mu\,\exp(\gamma/2 +\varepsilon\zeta(2)/8)}{\sqrt{4\pi}},
\end{equation}

\noindent
where

\begin{equation}
\zeta(s) \equiv \sum\limits_{n=1}^\infty \frac{1}{n^s}
\end{equation}

\noindent
is the Riemann $\zeta$-function. Note that in the four-loop approximation RGFs for the model (\ref{Varphi4_Lagrangian}) were obtained in \cite{Kazakov:1979ik}. The five-loop anomalous dimension of the field $\varphi$ was obtained in \cite{Chetyrkin:1981jq}, where one of the diagrams was calculated numerically. The complete analytic expression for the five-loop anomalous dimension of the field $\varphi$ can be obtained using the result presented in \cite{Chetyrkin:1981qh}. The five-loop $\beta$-function and the mass anomalous dimension were found in \cite{Gorishnii:1983gp}, but three of 124 diagrams were not calculated analytically. The analytical calculation of the five-loop $\beta$-function was completed in \cite{Kazakov:1983dyk,Kazakov:1983ns}. After some corrections the final five-loop results were presented in \cite{Kleinert:1991rg}. The six-loop RGFs for the $\varphi^4$ model can be found in \cite{Kompaniets:2017yct}. For the general scalar theory they have been calculated in \cite{Bednyakov:2021ojn}. Various recursion relations for the renormalization constants in higher orders have been verified in \cite{Kleinert:2001ax,Kastening:1996nj,Kastening:1997ah}. Here we use the expressions for various renormalization constants for checking the general results derived above.

The charge renormalization constant $Z_g$ is the analog of the renormalization constant $Z_\alpha$. (However, we reserve the letter $\alpha$ for the gauge coupling constant and use here the notation $Z_g$.) According to Eq. (15.16) of \cite{Kleinert:2001ax},

\begin{eqnarray}\label{Z_G_Varphi4}
&&	(Z_g)^{-1} = 1 + g \frac{(8 + N)}{3 \varepsilon} + g^2 \bigg[ - \frac{(14 + 3N)}{6 \varepsilon } + \frac{(8 + N)^2}{9\varepsilon^2} \bigg]\nonumber\\
&& + g^3 \bigg[ \frac{1}{648 \varepsilon} \Big( 2960 + 922N + 33N^2 + \zeta(3) (2112 + 480N) \Big) - \frac{7}{54 \varepsilon^2 } (8 + N) (14 + 3N)\nonumber\\
&&\qquad + \frac{(8 + N)^3}{27 \varepsilon^3} \bigg]
\nonumber \\	
&& + g^4 \bigg[ \frac{1}{15552 \varepsilon} \Big(-196648 - 80456N - 6320 N^2 + 5N^3
\nonumber\\
&&\qquad
- \zeta(3)(223872 + 73344N + 6048N^2)
\vphantom{\frac{1}{2}}
\nonumber\\
&&\qquad + \zeta(4)(50688 + 17856N + 1440N^2)
\vphantom{\frac{1}{2}}
\nonumber\\
&&\qquad
- \zeta(5) (357120 + 105600N + 3840N^2) \Big)
\vphantom{\frac{1}{2}}
\nonumber\\
&&\qquad + \frac{1}{3888 \varepsilon^2} \Big(150152 + 65288N + 7388N^2 + 165N^3
\nonumber\\
&&\qquad + \zeta(3) (84480 + 29760N + 2400N^2)  \Big)
\vphantom{\frac{1}{2}}
\nonumber\\
&&\qquad - \frac{23}{324 \varepsilon^3} (8 + N)^2 (14 + 3N) +\frac{(8 + N)^4}{81 \varepsilon^4} \bigg]
\nonumber\\
&&
+ g^5 \bigg[ \frac{1}{311040 \varepsilon} \Big( 13177344 + 6646336N + 808496N^2 + 12578N^3 + 13N^4
\nonumber \\	
&&\qquad + \zeta(3) (21029376+8836480N+1082240N^2+19968N^3-144N^4)
\vphantom{\frac{1}{2}}
\nonumber \\	
&&\qquad + \zeta^2(3) (2506752 + 342528N - 45312N^2 - 4608N^3)
\vphantom{\frac{1}{2}}
\nonumber \\	
&&\qquad - \zeta(4)(6082560 + 2745216N + 399744N^2 + 18144N^3)
\vphantom{\frac{1}{2}}
\nonumber \\	
&&\qquad + \zeta(5) (42261504 + 17148416N + 1911296N^2 + 78080N^3)
\vphantom{\frac{1}{2}}
\nonumber \\	
&&\qquad - \zeta(6) (14284800 + 6009600N + 681600N^2 + 19200N^3)
\vphantom{\frac{1}{2}}
\nonumber \\	
&&\qquad + \zeta(7) (59383296 + 21337344N + 1580544N^2) \Big)
\vphantom{\frac{1}{2}}
\nonumber\\
&&\qquad + \frac{1}{233280 \varepsilon^2 } \Big( -28905152 - 15368600N - 2361720N^2 - 101836N^3 + 65N^4
\nonumber \\	
&&
\qquad - \zeta(3) (29314560 + 13201536N + 1876224N^2 + 78624N^3)
\vphantom{\frac{1}{2}}
\nonumber\\	
&&\qquad + \zeta(4) (5271552 + 2515968N + 381888N^2 + 18720N^3)
\vphantom{\frac{1}{2}}
\nonumber \\	
&&\qquad -\zeta(5) (37140480+15624960N+1772160N^2+49920N^3) \Big)
\vphantom{\frac{1}{2}}
\nonumber\\
&& \qquad
+ \frac{(8 + N)}{58320 \varepsilon^3 } \Big( 1572136 + 681832N + 76432N^2 + 1419N^3
\nonumber\\
&&\qquad + \zeta(3) (726528 + 255936N + 20640N^2) \Big)
\vphantom{\frac{1}{2}}
\nonumber \\
&&\qquad - \frac{163}{4860 \varepsilon^4}(8 + N)^3 (14 + 3N)
+\frac{(8 + N)^5}{243 \varepsilon^5}
\bigg]+ O(g^6).
\end{eqnarray}

\noindent
For $S=-1$ the coefficients at $g^L/\varepsilon$ are equal to $\beta_L/L$, where $\beta_L$ are the coefficients of the $\beta$-function. Therefore, we conclude that

\begin{eqnarray}\label{Beta_G_Lowest}
&&\hspace*{-4mm} \beta_1 = \frac{(8 + N)}{3};\qquad \beta_2 =  - \frac{(14 + 3N)}{3};
\nonumber\\
&&\hspace*{-4mm} \beta_3 = \frac{1}{216} \Big( 2960 + 922N + 33N^2 + \zeta(3) (2112 + 480N) \Big);
\nonumber\\
&&\hspace*{-4mm} \beta_4 = \frac{1}{3888} \Big(-196648 - 80456N - 6320 N^2 + 5N^3
- \zeta(3)(223872 + 73344N + 6048N^2)
\nonumber\\
&&\hspace*{-4mm} \qquad + \zeta(4)(50688 + 17856N + 1440N^2) - \zeta(5) (357120 + 105600N + 3840N^2) \Big).
\vphantom{\frac{1}{2}}
\end{eqnarray}

\noindent
Note that we do not present the (rather large) expression for $\beta_5$ because it is not needed for calculating the coefficients at higher poles in the considered (five-loop) approximation. Now it is possible to compare the expression (\ref{Z_G_Varphi4}) with the prediction of Eq. (\ref{ZS_5Loops}). Extracting the terms with pure $\varepsilon$-poles, setting $S=-1$, and replacing $\alpha$ by $g$ we obtain that $(Z_g)^{-1}$ should have the structure
\footnote{Up to notation, this expression agrees with the two-loop expression presented in \cite{Kastening:1996nj}.}

\begin{eqnarray}\label{Z_S=-1_5Loops}
&&\hspace*{-7mm}  (Z_g)^{-1} = 1 + \frac{g \beta_1}{\varepsilon} + g^2 \Big( \frac{\beta_2}{2\varepsilon}  +  \frac{\beta_1^2}{\varepsilon^2} \Big)
+ g^3 \Big( \frac{\beta_3}{3\varepsilon} + \frac{7\beta_1\beta_2}{6\varepsilon^2} + \frac{\beta_1^3}{\varepsilon^3} \Big)
+ g^4 \Big(\frac{\beta_4}{4\varepsilon} +\frac{20\beta_1\beta_3 + 9\beta_2^2}{24\varepsilon^2}
\nonumber\\
&& \hspace*{-7mm} + \frac{23\beta_1^2\beta_2}{12\varepsilon^3} + \frac{\beta_1^4}{\varepsilon^4} \Big)
+ g^5 \Big( \frac{\beta_5}{5\varepsilon} + \frac{39\beta_1\beta_4 + 34\beta_2\beta_3}{60\varepsilon^2}
+ \frac{172\beta_1^2\beta_3 + 157\beta_1\beta_2^2}{120\varepsilon^3} + \frac{163\beta_1^3\beta_2}{60\varepsilon^4} + \frac{\beta_1^5}{\varepsilon^5} \Big)\nonumber\\
&&\hspace*{-7mm} + O(g^6). \vphantom{\frac{1}{\varepsilon^2}}
\end{eqnarray}

\noindent
We have substituted into this expression the coefficients of the $\beta$-function given by Eq. (\ref{Beta_G_Lowest}). The result exactly coincided with Eq. (\ref{Z_G_Varphi4}). Therefore, this calculation confirms the correctness of the general result (\ref{ZS_Final_Expansion}) derived in Section \ref{Section_Z_Powers}. Note that making this verification we were not checking the 't~Hooft pole equations, but their solutions.

It is also possible to verify the expressions for $Z_\varphi$ and $Z_m$ using a similar method. In particular, the five-loop expression for $Z_\varphi$ is given by Eq. (15.11) in \cite{Kleinert:2001ax},

\begin{eqnarray}\label{Z_Varphi_Varphi4}
&&\hspace*{-2mm} Z_\varphi = 1 - g^2 \frac{(2+N)}{36\varepsilon} + g^3 (2+N)(8+N)\bigg[\frac{1}{648\varepsilon} - \frac{1}{162\varepsilon^2}\bigg]
\nonumber\\
&&\hspace*{-2mm} + g^4 (2+N) \bigg[ \frac{5}{10368\varepsilon}\Big(-100 - 18 N + N^2 \Big) + \frac{1}{2592\varepsilon^2}\Big(234 + 53 N + N^2\Big) -\frac{(8+N)^2}{648\varepsilon^3} \bigg]
\nonumber\\
&&\hspace*{-2mm} + g^5 (2+N) \bigg[
\frac{1}{466560\varepsilon}\Big(77056 + 22752 N + 296 N^2 + 39 N^3\nonumber\\
&&\hspace*{-2mm}\qquad - \zeta(3) (8832+3072 N-288N^2+48 N^3) + \zeta(4)(25344 + 5760 N)\Big)
\nonumber\\
&&\hspace*{-2mm}\qquad + \frac{1}{116640\varepsilon^2}\Big(-33872-10610 N - 461 N^2 + 15 N^3 - \zeta(3)(12672 + 2880 N)\Big)
\nonumber\\
&&\hspace*{-2mm}\qquad + \frac{1}{29160\varepsilon^3} (8+N)\Big(1210+269 N +3N^2\Big)
- \frac{(8+N)^3}{2430\varepsilon^4} \bigg] + O(g^6).
\end{eqnarray}

\noindent
In this paper we present the general result only for the logarithm of the field renormalization constant, so that, first, it is necessary to calculate the logarithm of the expression (\ref{Z_Varphi_Varphi4}). In the considered (five-loop) approximation it is written as

\begin{eqnarray}\label{Z_Varphi_Logarithm}
&&\hspace*{-1mm} \ln Z_\varphi = - g^2 \frac{(2+N)}{36\varepsilon} + g^3 (2+N)(8+N)\bigg[\frac{1}{648\varepsilon} - \frac{1}{162\varepsilon^2}\bigg]
\nonumber\\
&&\hspace*{-1mm} + g^4 (2+N) \bigg[ \frac{5}{10368\varepsilon}\Big(-100 - 18 N + N^2 \Big) + \frac{1}{2592\varepsilon^2}\Big(232 + 52 N + N^2\Big) -\frac{(8+N)^2}{648\varepsilon^3} \bigg]
\nonumber\\
&&\hspace*{-1mm} + g^5 (2+N) \bigg[
\frac{1}{466560\varepsilon}\Big(77056 + 22752 N + 296 N^2 + 39 N^3\nonumber\\
&&\hspace*{-1mm}\qquad - \zeta(3) (8832+3072 N-288N^2+48 N^3) + \zeta(4)(25344 + 5760 N)\Big)
\nonumber\\
&&\hspace*{-1mm}\qquad + \frac{1}{116640\varepsilon^2}\Big(-33792-10560 N - 456 N^2 + 15 N^3 - \zeta(3)(12672 + 2880 N)\Big)
\nonumber\\
&&\hspace*{-1mm}\qquad + \frac{1}{29160\varepsilon^3} (8+N)\Big(1200+264 N +3N^2\Big)
- \frac{(8+N)^3}{2430\varepsilon^4} \bigg] + O(g^6).
\end{eqnarray}

\noindent
Considering the terms of the order $1/\varepsilon$ we see that the coefficients in the anomalous dimension of the field $\varphi$ up to the five-loop approximation are given by the expressions

\begin{eqnarray}
&& (\gamma_\varphi)_1 = 0;\qquad (\gamma_\varphi)_2 = \frac{(2+N)}{18};\qquad (\gamma_\varphi)_3 = - \frac{(2+N)(8+N)}{216};\nonumber\\
&& (\gamma_\varphi)_4 = \frac{5(2+N)}{2592}\Big(100 + 18 N - N^2 \Big).
\end{eqnarray}

\noindent
Note that we again do not present the large expression for $(\gamma_\varphi)_5$ because it is not required for calculating the coefficients at higher poles. To verify the general results for the higher poles in the expression for $\ln Z$ derived above, we should compare the coefficients at higher $\varepsilon$-poles in Eq. (\ref{Z_Varphi_Logarithm}) with the expression (\ref{LnZM_5Loops_Poles}) following from Eqs. (\ref{lnZM_Final_Expansion}) and (\ref{LnZM_Final_Result}). Having substituted the above values of the coefficients $(\gamma_\varphi)_L$ and $\beta_L$ into Eq. (\ref{LnZM_5Loops_Poles}) we obtained exactly the expression (\ref{Z_Varphi_Logarithm}) thus confirming the correctness of the results presented in Section \ref{Section_Field_Renormalization}.

The result for the renormalization constant $Z_m$ (defined by Eq. (\ref{Varphi4_Renormalization_Constants})) can also be found in \cite{Kleinert:2001ax}, where it is given by Eq. (15.15),\footnote{The general results obtained in Section~\ref{Section_Field_Renormalization} are also valid for the mass renormalization.}

\begin{eqnarray}\label{Z_Varphi4_M_5Loops}
&& Z_m = 1 + \frac{g (2 + N)}{3\varepsilon}
+ g^2 (2 + N)\bigg[ - \frac{5}{36 \varepsilon} + \frac{(5 + N)}{9\varepsilon^2} \bigg]
\nonumber\\
&& + g^3 (2 + N)
\bigg[ \frac{1}{108 \varepsilon} (37 + 5 N)
-\frac{1}{324\varepsilon^2} (278 + 61 N)
+ \frac{(5 + N) (6 + N)}{27\varepsilon^3}
\bigg]
\nonumber\\
&& + g^4 (2 + N) \bigg[
\frac{1}{31104\varepsilon} \Big(-31060 -7578 N + N^2
\nonumber\\
&&\qquad -\zeta(3) (3264 + 480 N + 144 N^2) - \zeta(4) (6336 + 1440 N)\Big)
\vphantom{\frac{1}{2}}
\nonumber\\
&&\qquad
+ \frac{1}{2592 \varepsilon^2} \Big(6218 + 1965 N + 103 N^2 + \zeta(3) (2112 + 480 N) \Big)
\nonumber\\
&&\qquad - \frac{1}{1944\varepsilon^3} \Big(6284 + 2498 N + 245 N^2 \Big)
+ \frac{1}{162 \varepsilon^4} (5 + N) (6 + N) (13 + 2 N)
\bigg]
\qquad\nonumber\\
&& + g^5 (2 + N)
\bigg[ \frac{1}{933120 \varepsilon} \Big(3166528 + 1077120 N + 45254 N^2 + 21 N^3
\nonumber\\
&&\qquad + \zeta(3) (1528704 + 393984 N + 45120 N^2 + 816 N^3)
\vphantom{\frac{1}{2}}
\nonumber\\
&&\qquad - \zeta(3)^2 (446976 + 111360 N + 1536 N^2)
\vphantom{\frac{1}{2}}
\nonumber\\
&&\qquad + \zeta(4) (768384 + 235008 N + 8352 N^2 -864 N^3)
\vphantom{\frac{1}{2}}
\nonumber\\
&&\qquad + \zeta(5) (55296 + 10752 N - 3840 N^2)
\vphantom{\frac{1}{2}}
\nonumber\\
&&\qquad + \zeta(6) (1785600 + 528000 N + 19200 N^2)\Big)
\vphantom{\frac{1}{2}}
\nonumber\\
&&\qquad + \frac{1}{466560 \varepsilon^2} \Big(-3724856 -1536688 N- 138640 N^2 + 49 N^3
\vphantom{\frac{1}{2}}
\nonumber\\
&&\qquad -\zeta(3) (2181504 + 693888 N + 58752 N^2 + 1296 N^3)
\vphantom{\frac{1}{2}}
\nonumber\\
&&\qquad
+ \zeta(4) (139392 + 25344 N -1440 N^2)
\vphantom{\frac{1}{2}}
\nonumber\\
&&\qquad - \zeta(5) (2856960 + 844800 N + 30720 N^2) \Big)
\vphantom{\frac{1}{2}}
\nonumber\\
&&\qquad + \frac{1}{116640 \varepsilon^3} \Big(1307420 + 627164 N + 85649 N^2 + 2697 N^3
\nonumber\\
&&\qquad + \zeta(3) (468864 + 188928 N + 18720 N^2) \Big)
\vphantom{\frac{1}{2}}
\nonumber\\
&&\qquad -\frac{1}{29160\varepsilon^4} (307976 + 172176 N + 31752 N^2 + 1933 N^3)
\nonumber\\
&&\qquad +\frac{1}{2430 \varepsilon^5} (5 + N) (6 + N) (13 + 2 N) (34 + 5 N)
\bigg] + O(g^6).
\end{eqnarray}

\noindent
Taking into account that the coefficients at $g^L/\varepsilon$ in this expression are equal to $-(\gamma_m)_L/L$, where $(\gamma_m)_L$ is the $L$-loop contribution to the mass anomalous dimension\footnote{The mass anomalous dimension is defined in the same way as the field anomalous dimension, but the renormalization constant $Z$ should in this case be replaced by $Z_m$.}, we conclude that

\begin{eqnarray}
&& (\gamma_m)_1 = -\frac{(2 + N)}{3};\qquad
(\gamma_m)_2 = \frac{5(2 + N)}{18};\qquad
(\gamma_m)_3 = - \frac{1}{36} (2 + N) (37 + 5 N);\qquad
\nonumber\\
&& (\gamma_m)_4 =  - \frac{(2 + N)}{7776} \Big(-31060 -7578 N + N^2
-\zeta(3) (3264 + 480 N + 144 N^2)
\nonumber\\
&&\qquad - \zeta(4) (6336 + 1440 N)\Big).\qquad
\end{eqnarray}

\noindent
Again we do not write down the large expression for $(\gamma_m)_5$ because it does not enter into the expressions for coefficients at higher poles in the considered (five-loop) approximation. We have calculated the logarithm of the expression (\ref{Z_Varphi4_M_5Loops}) and compared the result (\ref{Z_Varphi4_M_Logarithm_5Loops}) presented in Appendix \ref{Appendix_Z_Varphi_M_Logarithm} with Eq. (\ref{LnZM_5Loops_Poles}). They exactly coincide. Therefore, all five-loop expressions for the renormalization constants of the theory (\ref{Varphi4_Lagrangian}) agree with the general expressions obtained in this paper.

\section{Conclusion}
\hspace*{\parindent}

In this paper we investigated the structure of the renormalization constants for such a version of the dimensional technique in that the dimensionful regularization parameter $\Lambda$ does not coincide with the renormalization point $\mu$. In this case in addition to $\varepsilon$-poles the renormalization constants also contain powers of $\ln\Lambda/\mu$ and the mixed terms. We have constructed the explicit all-loop expressions which relate all coefficients at higher $\varepsilon$-poles, logarithms, and mixed terms to the coefficients of RGFs (i.e., of the $\beta$-function and of the anomalous dimension). These equations have been written for $\ln Z_\alpha$, $(Z_\alpha)^S$, and $\ln Z$, where $Z_\alpha$ and $Z$ are the charge and field renormalization constants. The general results are given by Eqs. (\ref{lnZ_Final_Expansion}), (\ref{ZS_Final_Expansion}), and (\ref{lnZM_Final_Expansion}), respectively. They can also be rewritten as the all-loop equations (\ref{LnZ_Final_Result}), (\ref{ZS_Final_Result}), and (\ref{LnZM_Final_Result}). In the lowest loops we present the explicit expressions following from these general equations. They have been verified by comparing with the results of some previous calculations. We have also revealed some interesting features of the general results. In particular, we explain how one can transform the result containing $\varepsilon$-poles obtained in the standard $\mbox{MS}$ ($\overline{\mbox{MS}}$, $\mbox{DR}$, $\overline{\mbox{DR}}$, etc.) schemes into the result containing pure logarithms (which appears, e.g., in the HD+MSL scheme). Certainly, we discussed only the dependence of the renormalization constants on the coefficients in RGFs and did not discuss how the coefficients of the renormalization group functions depend on a renormalization scheme.

\section*{Acknowledgments}
\hspace*{\parindent}

K.S. is very grateful to A.L.Kataev and D.I.Kazakov for indicating the problem addressed in this paper.

This work has been supported by Foundation for Advancement of Theoretical Physics and Mathematics ``BASIS'', grant  No. 21-2-2-25-1 (N.M.) and by Russian Science Foundation, grant No. 21-12-00129 (K.S.).

\appendix

\section{Explicit expressions for various functions of the renormalization constants in the lowest loops}
\hspace*{\parindent}\label{Appendix_Explicit_Z}

In this appendix we present the explicit five-loop expressions for $\ln Z_\alpha$, $(Z_\alpha)^S$, and $\ln Z$ and the explicit six-loop expression for $Z_\alpha$ calculated with the help of Eqs. (\ref{lnZ_Final_Expansion}), (\ref{ZS_Final_Expansion}), and (\ref{lnZM_Final_Expansion}).

The five-loop expression for $\ln Z_\alpha$ is

\begin{eqnarray}\label{LnZ_5Loops}
&&  \ln Z_\alpha = - \alpha \beta_1 \Big(\frac{1}{\varepsilon}+ \ln\frac{\Lambda}{\mu} \Big)
-  \frac{\alpha^2}{2} \bigg[ \beta_2 \Big(\frac{1}{\varepsilon}+ 2 \ln\frac{\Lambda}{\mu} \Big)+ \beta_1^2 \Big(\frac{1}{\varepsilon}+ \ln \frac{\Lambda}{\mu} \Big)^2  \bigg]
\nonumber\\
&& - \frac{\alpha^3}{3} \bigg[ \beta_3 \Big(\frac{1}{\varepsilon}+ 3 \ln \frac{\Lambda}{\mu} \Big) + 2 \beta_1\beta_2 \Big(\frac{1}{\varepsilon}+ \frac{3}{2}\ln \frac{\Lambda}{\mu} \Big)^2
+ \beta_1^3\Big(\frac{1}{\varepsilon}+ \ln \frac{\Lambda}{\mu} \Big)^3\bigg]
\nonumber\\
&& - \frac{\alpha^4}{4} \bigg[ \beta_4 \Big(\frac{1}{\varepsilon}+ 4 \ln \frac{\Lambda}{\mu}\Big)
+\big(2\beta_1\beta_3  +\beta_2^2\big) \Big(\frac{1}{\varepsilon}+ 2 \ln\frac{\Lambda}{\mu} \Big)^2
\nonumber\\
&&\qquad + 3 \beta_1^2\beta_2 \Big(\frac{1}{\varepsilon^3} + \frac{4}{\varepsilon^2} \ln\frac{\Lambda}{\mu} + \frac{16}{3 \varepsilon} \ln^2\frac{\Lambda}{\mu} + \frac{22}{9} \ln^3\frac{\Lambda}{\mu} \Big)
+ \beta_1^4\Big(\frac{1}{\varepsilon}+ \ln \frac{\Lambda}{\mu} \Big)^4  \bigg]
\nonumber\\
&& - \frac{\alpha^5}{5} \bigg[ \beta_5 \Big(\frac{1}{\varepsilon} + 5\ln\frac{\Lambda}{\mu}\Big) + 2(\beta_1\beta_4+\beta_2\beta_3) \Big(\frac{1}{\varepsilon} +\frac{5}{2}\ln\frac{\Lambda}{\mu}\Big)^2
+ \beta_1^5\Big(\frac{1}{\varepsilon} + \ln\frac{\Lambda}{\mu} \Big)^5
\nonumber\\
&&\qquad + 3 \beta_1^2 \beta_3 \Big(\frac{1}{\varepsilon^3}+ \frac{5}{\varepsilon^2} \ln\frac{\Lambda}{\mu} + \frac{25}{3\varepsilon} \ln^2\frac{\Lambda}{\mu} + 5\ln^3\frac{\Lambda}{\mu} \Big)
\nonumber\\
&&\qquad + 3\beta_1\beta_2^2\Big(\frac{1}{\varepsilon^3}+ \frac{5}{\varepsilon^2} \ln \frac{\Lambda}{\mu} + \frac{25}{3 \varepsilon} \ln^2\frac{\Lambda}{\mu} + \frac{85}{18} \ln^3\frac{\Lambda}{\mu} \Big)
\nonumber\\
&&\qquad + 4\beta_1^3\beta_2 \Big(\frac{1}{\varepsilon^4}+ \frac{5}{\varepsilon^3} \ln \frac{\Lambda}{\mu}+ \frac{75}{8\varepsilon^2} \ln^2\frac{\Lambda}{\mu}
+ \frac{95}{12 \varepsilon} \ln^3\frac{\Lambda}{\mu} + \frac{125}{48} \ln^4\frac{\Lambda}{\mu} \Big) \bigg] + O(\alpha^6).\qquad
\end{eqnarray}

\noindent
The five-loop expression for $(Z_\alpha)^S/S$ is

\begin{eqnarray}\label{ZS_5Loops}
&&  \frac{(Z_\alpha)^S}{S} = \frac{1}{S} - \alpha \beta_1 \Big(\frac{1}{\varepsilon}+ \ln \frac{\Lambda}{\mu} \Big)
- \alpha^2 \bigg[ \frac{\beta_2}{2} \Big(\frac{1}{\varepsilon}+ 2 \ln \frac{\Lambda}{\mu} \Big) +  \frac{\beta_1^2}{2} (1-S) \Big(\frac{1}{\varepsilon} + \ln\frac{\Lambda}{\mu} \Big)^2 \bigg]
\nonumber\\
&& - \alpha^3 \bigg[ \frac{\beta_3}{3} \Big(\frac{1}{\varepsilon} + 3\ln\frac{\Lambda}{\mu} \Big)
+ \frac{ \beta_1\beta_2}{6} \bigg\{(4-3S) \Big(\frac{1}{\varepsilon^2} + \frac{3}{\varepsilon} \ln\frac{\Lambda}{\mu}\Big)+ 3 (3-2S) \ln^2\frac{\Lambda}{\mu} \bigg\}
\nonumber\\
&&\qquad + \frac{\beta_1^3}{6} (1-S)(2-S) \Big(\frac{1}{\varepsilon} + \ln\frac{\Lambda}{\mu} \Big)^3 \bigg]
\nonumber\\
&& - \alpha^4 \bigg[ \frac{\beta_4}{4} \Big(\frac{1}{\varepsilon}+ 4 \ln\frac{\Lambda}{\mu}\Big)
+\frac{ \beta_1\beta_3}{6} \bigg\{ (3-2S)\Big(\frac{1}{\varepsilon^2} + \frac{4}{\varepsilon} \ln\frac{\Lambda}{\mu}\Big) + 6(2-S) \ln^2\frac{\Lambda}{\mu} \bigg\}
\nonumber\\
&&\qquad + \frac{ \beta_2^2}{8}(2-S) \Big(\frac{1}{\varepsilon} + 2\ln\frac{\Lambda}{\mu} \Big)^2
+ \frac{\beta_1^2\beta_2}{12} \bigg\{ \big(9-11S+3S^2\big) \Big(\frac{1}{\varepsilon^3}+ \frac{4}{\varepsilon^2} \ln \frac{\Lambda}{\mu}\Big)
\nonumber\\
&&\qquad + 3\big(16-19S+5S^2\big) \frac{1}{\varepsilon} \ln^2\frac{\Lambda}{\mu}+ 2\big(11-12S+3S^2\big)  \ln^3\frac{\Lambda}{\mu} \bigg\}
\nonumber\\
&&\qquad + \frac{\beta_1^4}{24}(1-S)(2-S)(3-S) \Big(\frac{1}{\varepsilon} + \ln\frac{\Lambda}{\mu} \Big)^4\bigg]
\nonumber\\
&& - \alpha^5 \bigg[ \frac{\beta_5}{5} \Big(\frac{1}{\varepsilon}+ 5 \ln \frac{\Lambda}{\mu} \Big)
+ \frac{\beta_1\beta_4}{20} \bigg\{ (8-5S)\Big(\frac{1}{\varepsilon^2} + \frac{5}{\varepsilon} \ln\frac{\Lambda}{\mu}\Big) + 10(5-2S)\ln^2\frac{\Lambda}{\mu} \bigg\}
\nonumber\\
&&\qquad + \frac{\beta_2\beta_3}{30} \bigg\{ (12-5S)\Big(\frac{1}{\varepsilon^2} + \frac{5}{\varepsilon} \ln\frac{\Lambda}{\mu}\Big)
+ 15(5-2S)\ln^2 \frac{\Lambda}{\mu} \bigg\}
\nonumber\\
&&\qquad + \frac{\beta_1^2\beta_3}{30} \bigg\{ \big(18-20S+5S^2\big)\Big(\frac{1}{\varepsilon^3}+ \frac{5}{\varepsilon^2} \ln \frac{\Lambda}{\mu}\Big)
+5\big(30-31S+7S^2\big) \frac{1}{\varepsilon} \ln^2 \frac{\Lambda}{\mu} \nonumber\\
&&\qquad + 15\big(6-5S+S^2\big)  \ln^3\frac{\Lambda}{\mu} \bigg\}
+ \frac{ \beta_1\beta_2^2}{120} \bigg\{\big(72-70S+15S^2\big) \Big(\frac{1}{\varepsilon^3}+ \frac{5}{\varepsilon^2} \ln \frac{\Lambda}{\mu}\Big)
\nonumber\\
&&\qquad +30\big(20-19S+4S^2\big) \frac{1}{ \varepsilon} \ln^2\frac{\Lambda}{\mu} + 20 \big(17-15S+3S^2\big) \ln^3\frac{\Lambda}{\mu} \bigg\}
\nonumber\\
&&\qquad + \frac{\beta_1^3\beta_2}{12} \bigg\{\frac{1}{5}\big(48-75S+35S^2-5S^3\big)\Big(\frac{1}{\varepsilon^4}+ \frac{5}{\varepsilon^3} \ln\frac{\Lambda}{\mu}\Big)\nonumber\\
&&\qquad + \big(90-139S+64S^2-9S^3\big)\frac{1}{\varepsilon^2} \ln^2 \frac{\Lambda}{\mu}
+ \big(76-114S+51S^2-7S^3\big) \frac{1}{\varepsilon} \ln^3\frac{\Lambda}{\mu}
\qquad\nonumber\\
&&\qquad + \big(25-35S+15S^2-2S^3\big) \ln^4\frac{\Lambda}{\mu} \bigg\}
\nonumber\\
&&\qquad + \frac{\beta_1^5}{120}(1-S)(2-S)(3-S)(4-S) \Big(\frac{1}{\varepsilon} + \ln \frac{\Lambda}{\mu}  \Big)^5
\bigg] + O(\alpha^6).
\end{eqnarray}

\noindent
For $S=1$ the six-loop expression for $Z_\alpha$ is written below. In the terms proportional to $(\beta_1)^k \beta_{L-k}$ (which are described by Eq. (\ref{Z_Alpha_Main_Sequence})) the coefficients of the $\beta$-function are indicated by the bold font.

\begin{eqnarray}\label{Z_Alpha_6Loops}
&&  Z_\alpha = 1 - \alpha \bm{\beta_1} \Big(\frac{1}{\varepsilon}+ \ln \frac{\Lambda}{\mu} \Big)
- \alpha^2 \frac{\bm{\beta_2}}{2} \Big(\frac{1}{\varepsilon}+ 2 \ln \frac{\Lambda}{\mu} \Big)
\nonumber\\
&& - \alpha^3 \bigg[ \frac{\bm{\beta_3}}{3} \Big(\frac{1}{\varepsilon}+ 3 \ln \frac{\Lambda}{\mu} \Big) +\frac{\bm{\beta_1\beta_2}}{6} \Big(\frac{1}{\varepsilon^2}+ \frac{3}{\varepsilon} \ln \frac{\Lambda}{\mu}
+ 3 \ln^2 \frac{\Lambda}{\mu} \Big)
\bigg]
\nonumber\\
&& - \alpha^4 \bigg[
\frac{\bm{\beta_4}}{4} \Big(\frac{1}{\varepsilon}+ 4 \ln\frac{\Lambda}{\mu}\Big)
+ \frac{\bm{\beta_1\beta_3}}{6} \Big(\frac{1}{\varepsilon^2}+ \frac{4}{\varepsilon} \ln\frac{\Lambda}{\mu} + 6\ln^2\frac{\Lambda}{\mu} \Big)
\nonumber\\
&&\qquad + \frac{\bm{\beta_1^2\beta_2}}{12} \Big(\frac{1}{\varepsilon^3}+ \frac{4}{\varepsilon^2} \ln \frac{\Lambda}{\mu} + \frac{6}{\varepsilon} \ln^2 \frac{\Lambda}{\mu} + 4 \ln^3\frac{\Lambda}{\mu} \Big)
+ \frac{\beta_2^2}{8} \Big(\frac{1}{\varepsilon^2}+ \frac{4}{\varepsilon} \ln \frac{\Lambda}{\mu} + 4 \ln^2 \frac{\Lambda}{\mu}  \Big)
\bigg]
\qquad\nonumber\\ 	
&& - \alpha^5 \bigg[
\frac{\bm{\beta_5}}{5} \Big(\frac{1}{\varepsilon}+ 5 \ln \frac{\Lambda}{\mu} \Big)
+ \frac{3\bm{\beta_1\beta_4}}{20} \Big(\frac{1}{\varepsilon^2} + \frac{5}{\varepsilon}\ln\frac{\Lambda}{\mu} + 10\ln^2\frac{\Lambda}{\mu} \Big)
\nonumber\\
&&\qquad + \frac{\bm{\beta_1^2\beta_3}}{10} \Big(\frac{1}{\varepsilon^3}+ \frac{5}{\varepsilon^2} \ln\frac{\Lambda}{\mu} + \frac{10}{\varepsilon}\ln^2\frac{\Lambda}{\mu} + 10  \ln^3 \frac{\Lambda}{\mu} \Big)
\nonumber\\
&&\qquad + \frac{\bm{\beta_1^3\beta_2}}{20} \Big(\frac{1}{\varepsilon^4}+ \frac{5}{\varepsilon^3} \ln\frac{\Lambda}{\mu}
+ \frac{10}{\varepsilon^2} \ln^2\frac{\Lambda}{\mu}	+ \frac{10}{\varepsilon} \ln^3\frac{\Lambda}{\mu} + 5\ln^4\frac{\Lambda}{\mu} \Big)
\nonumber\\
&&\qquad + \frac{7\beta_2\beta_3}{30}\Big(\frac{1}{\varepsilon^2}+ \frac{5}{\varepsilon} \ln\frac{\Lambda}{\mu} + \frac{45}{7}\ln^2\frac{\Lambda}{\mu} \Big)
\nonumber\\
&&\qquad + \frac{ 17 \beta_1\beta_2^2}{120} \Big(\frac{1}{\varepsilon^3}+ \frac{5}{\varepsilon^2} \ln \frac{\Lambda}{\mu}
+ \frac{150}{17 \varepsilon} \ln^2 \frac{\Lambda}{\mu}
+ \frac{100}{17}  \ln^3 \frac{\Lambda}{\mu} \Big)
\bigg]
\nonumber\\ 	
&& - \alpha^6 \bigg[
\frac{\bm{\beta_6}}{6} \Big(\frac{1}{\varepsilon}+ 6 \ln \frac{\Lambda}{\mu} \Big)
+ \frac{2\bm{\beta_1\beta_5}}{15} \Big(\frac{1}{\varepsilon^2}+ \frac{6}{\varepsilon} \ln \frac{\Lambda}{\mu} + 15 \ln^2 \frac{\Lambda}{\mu}  \Big)
\nonumber\\
&&\qquad + \frac{\bm{\beta_1^2\beta_4}}{10} \Big(\frac{1}{\varepsilon^3}+ \frac{6}{\varepsilon^2} \ln\frac{\Lambda}{\mu}
+ \frac{15}{\varepsilon}\ln^2\frac{\Lambda}{\mu} + 20\ln^3\frac{\Lambda}{\mu}  \Big)
\nonumber\\
&&\qquad + \frac{\bm{\beta_1^3\beta_3}}{15} \Big(\frac{1}{\varepsilon^4}+ \frac{6}{\varepsilon^3} \ln\frac{\Lambda}{\mu}	+ \frac{15}{\varepsilon^2} \ln^2\frac{\Lambda}{\mu}
+ \frac{20}{\varepsilon} \ln^3\frac{\Lambda}{\mu} + 15\ln^4\frac{\Lambda}{\mu} \Big)
\nonumber\\
&&\qquad + \frac{\bm{\beta_1^4\beta_2}}{30}\Big(\frac{1}{\varepsilon^5}+ \frac{6}{\varepsilon^4} \ln\frac{\Lambda}{\mu} + \frac{15}{\varepsilon^3} \ln^2 \frac{\Lambda}{\mu}
+ \frac{20}{\varepsilon^2}  \ln^3\frac{\Lambda}{\mu} + \frac{15}{\varepsilon} \ln^4\frac{\Lambda}{\mu} + 6 \ln^5\frac{\Lambda}{\mu} \Big)
\nonumber\\
&& \qquad +\frac{5\beta_2\beta_4}{24} \Big(\frac{1}{\varepsilon^2} + \frac{6}{\varepsilon} \ln\frac{\Lambda}{\mu} + \frac{48}{5} \ln^2\frac{\Lambda}{\mu} \Big)
+\frac{\beta_3^2}{9} \Big(\frac{1}{\varepsilon^2} + \frac{6}{\varepsilon} \ln\frac{\Lambda}{\mu} + 9\ln^2\frac{\Lambda}{\mu} \Big)
\nonumber\\
&&\qquad + \frac{53\beta_1\beta_2\beta_3}{180}\Big(\frac{1}{\varepsilon^3}+ \frac{6}{\varepsilon^2}\ln\frac{\Lambda}{\mu} + \frac{690}{53\varepsilon}\ln^2\frac{\Lambda}{\mu} + \frac{600}{53}\ln^3\frac{\Lambda}{\mu} \Big)
\nonumber\\
&&\qquad
+ \frac{49\beta_1^2\beta_2^2}{360} \Big(\frac{1}{\varepsilon^4} + \frac{6}{\varepsilon^3}\ln\frac{\Lambda}{\mu} + \frac{690}{49\varepsilon^2}\ln^2\frac{\Lambda}{\mu}
+ \frac{780}{49\varepsilon}\ln^3\frac{\Lambda}{\mu} + \frac{390}{49}\ln^4\frac{\Lambda}{\mu} \Big)
\nonumber\\
&&\qquad
+ \frac{\beta_2^3}{16} \Big(\frac{1}{\varepsilon^3} + \frac{6}{\varepsilon^2}\ln\frac{\Lambda}{\mu} + \frac{12}{\varepsilon}\ln^2\frac{\Lambda}{\mu} + 8\ln^3\frac{\Lambda}{\mu} \Big)
\bigg] + O(\alpha^7).
\end{eqnarray}

\noindent
The five-loop expression for $\ln Z$ is

\begin{eqnarray}\label{LnZM_5Loops}
&& \ln Z = - \alpha \gamma_1 \Big(\frac{1}{\varepsilon} + \ln\frac{\Lambda}{\mu} \Big) - \frac{\alpha^2}{2} \bigg[\gamma_2 \Big(\frac{1}{\varepsilon}+ 2 \ln\frac{\Lambda}{\mu} \Big)
+ \gamma_1\beta_1 \Big(\frac{1}{\varepsilon} + \ln\frac{\Lambda}{\mu} \Big)^2  \bigg]
\nonumber\\
&& - \frac{\alpha^3}{3} \bigg[ \gamma_3 \Big(\frac{1}{\varepsilon} + 3 \ln\frac{\Lambda}{\mu} \Big)
+ \gamma_1\beta_2 \Big(\frac{1}{\varepsilon^2} + \frac{3}{\varepsilon} \ln\frac{\Lambda}{\mu} + \frac{3}{2} \ln^2\frac{\Lambda}{\mu} \Big)
+ \gamma_2\beta_1 \Big(\frac{1}{\varepsilon^2} + \frac{3}{\varepsilon} \ln\frac{\Lambda}{\mu} + 3\ln^2\frac{\Lambda}{\mu} \Big)
\qquad\nonumber\\
&&\qquad + \gamma_1\beta_1^2 \Big(\frac{1}{\varepsilon} + \ln\frac{\Lambda}{\mu} \Big)^3\bigg]
\nonumber\\
&& - \frac{\alpha^4}{4} \bigg[ \gamma_4 \Big(\frac{1}{\varepsilon}+ 4 \ln\frac{\Lambda}{\mu}\Big)
+\gamma_1\beta_3 \Big(\frac{1}{\varepsilon^2} + \frac{4}{\varepsilon} \ln\frac{\Lambda}{\mu} + 2\ln^2\frac{\Lambda}{\mu} \Big)
+\gamma_2\beta_2 \Big(\frac{1}{\varepsilon} + 2\ln \frac{\Lambda}{\mu} \Big)^2
\nonumber\\
&&\qquad +\gamma_3\beta_1 \Big(\frac{1}{\varepsilon^2} + \frac{4}{\varepsilon} \ln\frac{\Lambda}{\mu} + 6\ln^2\frac{\Lambda}{\mu} \Big)
+ 2 \gamma_1\beta_1\beta_2 \Big(\frac{1}{\varepsilon^3} + \frac{4}{\varepsilon^2} \ln\frac{\Lambda}{\mu} + \frac{5}{\varepsilon} \ln^2\frac{\Lambda}{\mu} + \frac{5}{3} \ln^3\frac{\Lambda}{\mu} \Big)
\nonumber\\
&&\qquad + \gamma_2\beta_1^2 \Big(\frac{1}{\varepsilon^3} + \frac{4}{\varepsilon^2} \ln\frac{\Lambda}{\mu} + \frac{6}{\varepsilon} \ln^2\frac{\Lambda}{\mu} + 4\ln^3\frac{\Lambda}{\mu} \Big)
+ \gamma_1\beta_1^3 \Big(\frac{1}{\varepsilon} + \ln \frac{\Lambda}{\mu} \Big)^4 \bigg]
\nonumber\\
&& - \frac{\alpha^5}{5} \bigg[ \gamma_5 \Big(\frac{1}{\varepsilon}+ 5 \ln \frac{\Lambda}{\mu} \Big)
+ \gamma_1\beta_4 \Big(\frac{1}{\varepsilon^2}+ \frac{5}{\varepsilon} \ln\frac{\Lambda}{\mu} + \frac{5}{2} \ln^2\frac{\Lambda}{\mu} \Big)
\nonumber\\
&&\qquad + \gamma_2\beta_3 \Big(\frac{1}{\varepsilon^2} + \frac{5}{\varepsilon} \ln\frac{\Lambda}{\mu} + 5\ln^2\frac{\Lambda}{\mu} \Big)
+ \gamma_3\beta_2 \Big(\frac{1}{\varepsilon^2} + \frac{5}{\varepsilon} \ln\frac{\Lambda}{\mu} + \frac{15}{2} \ln^2 \frac{\Lambda}{\mu} \Big)
\nonumber\\
&&\qquad
+\gamma_2\beta_1^3 \Big(\frac{1}{\varepsilon^4} + \frac{5}{\varepsilon^3} \ln\frac{\Lambda}{\mu} + \frac{10}{\varepsilon^2} \ln^2\frac{\Lambda}{\mu} + \frac{10}{\varepsilon} \ln^3\frac{\Lambda}{\mu}
+ 5 \ln^4\frac{\Lambda}{\mu} \Big) +\gamma_1\beta_1^4 \Big(\frac{1}{\varepsilon} + \ln\frac{\Lambda}{\mu} \Big)^5
\nonumber\\
&&\qquad + \gamma_4\beta_1 \Big(\frac{1}{\varepsilon^2} + \frac{5}{\varepsilon} \ln\frac{\Lambda}{\mu} + 10\ln^2\frac{\Lambda}{\mu} \Big)
+ \gamma_3\beta_1^2 \Big(\frac{1}{\varepsilon^3} + \frac{5}{\varepsilon^2} \ln\frac{\Lambda}{\mu} + \frac{10}{\varepsilon} \ln^2\frac{\Lambda}{\mu} + 10\ln^3\frac{\Lambda}{\mu} \Big)
\nonumber\\
&&\qquad  + \big(2\gamma_1\beta_1\beta_3 + \gamma_1\beta_2^2\big) \Big(\frac{1}{\varepsilon^3} + \frac{5}{\varepsilon^2} \ln\frac{\Lambda}{\mu}	+ \frac{15}{2\varepsilon} \ln^2\frac{\Lambda}{\mu}
+ \frac{5}{2} \ln^3\frac{\Lambda}{\mu} \Big)
\nonumber\\
&&\qquad + 2 \gamma_2\beta_1\beta_2 \Big(\frac{1}{\varepsilon^3} + \frac{5}{\varepsilon^2} \ln\frac{\Lambda}{\mu} + \frac{35}{4 \varepsilon} \ln^2\frac{\Lambda}{\mu} + \frac{35}{6} \ln^3\frac{\Lambda}{\mu} \Big)
\nonumber\\
&&\qquad +  3\gamma_1\beta_1^2\beta_2 \Big(\frac{1}{\varepsilon^4} + \frac{5}{\varepsilon^3} \ln\frac{\Lambda}{\mu} + \frac{55}{6 \varepsilon^2} \ln^2\frac{\Lambda}{\mu}
+ \frac{65}{9\varepsilon} \ln^3\frac{\Lambda}{\mu} + \frac{65}{36} \ln^4\frac{\Lambda}{\mu} \Big)
\bigg] + O(\alpha^6).
\end{eqnarray}

\section{The logarithm of $Z_m$ for the $\varphi^4$-theory}
\hspace{\parindent}\label{Appendix_Z_Varphi_M_Logarithm}

To compare the mass renormalization constant for the $\varphi^4$-theory with the exact equation (\ref{lnZM_Final_Expansion}), first, it is necessary to calculate the logarithm of the expression (\ref{Z_Varphi4_M_5Loops}). The result is rather large and can be written as

\begin{eqnarray}\label{Z_Varphi4_M_Logarithm_5Loops}
&& \ln Z_m = \frac{g (2 + N)}{3\varepsilon}
+ g^2 (2 + N)\bigg[ - \frac{5}{36 \varepsilon} + \frac{(8 + N)}{18\varepsilon^2} \bigg]
\nonumber\\
&& + g^3 (2 + N)
\bigg[ \frac{1}{108 \varepsilon} (37 + 5 N)
-\frac{1}{162\varepsilon^2} (124 + 23 N)
+ \frac{(8+N)^2}{81\varepsilon^3}
\bigg]
\nonumber\\
&& + g^4 (2 + N) \bigg[
\frac{1}{31104\varepsilon} \Big(-31060 -7578 N + N^2
\nonumber\\
&&\qquad -\zeta(3) (3264 + 480 N + 144 N^2) - \zeta(4) (6336 + 1440 N)\Big)
\vphantom{\frac{1}{2}}
\nonumber\\
&&\qquad
+ \frac{1}{2592 \varepsilon^2} \Big(5576 + 1564 N + 63 N^2 + \zeta(3) (2112 + 480 N) \Big)
\nonumber\\
&&\qquad - \frac{1}{648\varepsilon^3} \Big(1664 + 536 N + 41 N^2 \Big)
+ \frac{(8 + N)^3}{324 \varepsilon^4} \bigg]
\qquad\nonumber\\
&& + g^5 (2 + N)
\bigg[ \frac{1}{933120 \varepsilon} \Big(3166528 + 1077120 N + 45254 N^2 + 21 N^3
\nonumber\\
&&\qquad + \zeta(3) (1528704 + 393984 N + 45120 N^2 + 816 N^3)
\vphantom{\frac{1}{2}}
\nonumber\\
&&\qquad - \zeta(3)^2 (446976 + 111360 N + 1536 N^2)
\vphantom{\frac{1}{2}}
\nonumber\\
&&\qquad + \zeta(4) (768384 + 235008 N + 8352 N^2 -864 N^3)
\vphantom{\frac{1}{2}}
\nonumber\\
&&\qquad + \zeta(5) (55296 + 10752 N - 3840 N^2)
\vphantom{\frac{1}{2}}
\nonumber\\
&&\qquad + \zeta(6) (1785600 + 528000 N + 19200 N^2)\Big)
\vphantom{\frac{1}{2}}
\nonumber\\
&&\qquad + \frac{1}{116640\varepsilon^2} \Big(-842464 - 319352 N- 24440 N^2 + 11 N^3
\vphantom{\frac{1}{2}}
\nonumber\\
&&\qquad -\zeta(3) (537216 + 168192 N + 13728 N^2 + 144 N^3)
\vphantom{\frac{1}{2}}
\nonumber\\
&&\qquad
+ \zeta(4) (50688 + 17856 N +1440 N^2)
\vphantom{\frac{1}{2}}
\nonumber\\
&&\qquad - \zeta(5) (714240 + 211200 N + 7680 N^2) \Big)
\vphantom{\frac{1}{2}}
\nonumber\\
&&\qquad + \frac{1}{4860 \varepsilon^3} \Big(44560 + 18376 N + 2041 N^2 + 48 N^3
\nonumber\\
&&\qquad + \zeta(3) (16896 + 5952 N + 480 N^2) \Big)
\vphantom{\frac{1}{2}}
\nonumber\\
&&\qquad -\frac{1}{2430\varepsilon^4} (18688 + 8448 N + 1236 N^2 + 59 N^3)
+\frac{(8 + N)^4}{1215\varepsilon^5}  \bigg] + O(g^6).
\end{eqnarray}

\end{document}